\documentclass[twocolumn,superscriptaddress,showpacs,nofootinbib,notitlepage,preprintnumbers,secnumarabic,amssymb, nobibnotes, aps, prd]{revtex4-2}
\usepackage[utf8]{inputenc}
\usepackage{graphicx}
\usepackage{latexsym,amsmath,amssymb,amsthm,lmodern,float,url}
\usepackage{natbib}
\usepackage{color}
\usepackage{microtype}
\usepackage{import}
\usepackage{bbold}
\usepackage[plain]{fancyref}
\usepackage{varioref}
\usepackage{slashed}
\usepackage{multirow}
\usepackage{tikz}
\usepackage{braket}
\usetikzlibrary{shapes}
\usetikzlibrary{positioning}
\usepackage[normalem]{ulem}

\definecolor{mycolor}{RGB}{22,139,22}

\newcommand{\blacksq}{\raisebox{0.5pt}{\tikz{\node[fill,scale=0.4,regular polygon, regular polygon sides=4,rotate=180](){};}}}
\newcommand{\blackosq}{\raisebox{0.5pt}{\tikz{\node[draw,scale=0.4,regular polygon, regular polygon sides=4,rotate=180](){};}}}

\newcommand{\plaq}{1\times1}
\newcommand{\tuplaq}{2\times1}

\newcommand{\fig}[1]{Fig.~\ref{fig:#1}}
\newcommand{\tab}[1]{Tab.~\ref{tab:#1}}
\newcommand{\eq}[1]{Eq.~(\ref{eq:#1})}

\usepackage[colorlinks=true,backref=false, linktocpage=true,
citecolor=blue,urlcolor=blue,linkcolor=blue,pdfpagemode=UseOutlines]{hyperref}
\hypersetup{%
  bookmarksnumbered=true,
  pdftitle = {},
  pdfsubject = {},
  pdfauthor = {},
  pdfkeywords = {}
}

\let\Re\undefined

\DeclareMathOperator{\Tr}{Tr}
\DeclareMathOperator{\Re}{Re}

\DeclareMathOperator{\tr}{Tr}
\DeclareMathOperator{\re}{Re}
\DeclareMathOperator{\im}{Im}
\newcommand{\oh}{\overline{H}}
\newtheorem{theorem}{Theorem}[section]
\newtheorem{lemma}[theorem]{Lemma}

\begin{document}
\preprint{FERMILAB-PUB-21-222-T}
\title{Lattice Renormalization of Quantum Simulations}
\author{Marcela Carena}
\email{carena@fnal.gov}
\affiliation{Fermi National Accelerator Laboratory, Batavia,  Illinois, 60510, USA}
\affiliation{Enrico Fermi Institute, University of Chicago, Chicago, Illinois, 60637, USA}
\affiliation{Kavli Institute for Cosmological Physics, University of Chicago, Chicago, Illinois, 60637, USA}
\author{Henry Lamm}
\email{hlamm@fnal.gov}
\affiliation{Fermi National Accelerator Laboratory, Batavia,  Illinois, 60510, USA}
\author{Ying-Ying Li}
\email{yingying@fnal.gov}
\affiliation{Fermi National Accelerator Laboratory, Batavia,  Illinois, 60510, USA}
\author{Wanqiang Liu}
\email{wanqiangl@uchicago.edu}
\affiliation{Department of Physics, University of Chicago, Chicago, Illinois, 60637, USA}
\begin{abstract}
With advances in quantum computing, new opportunities arise to tackle challenging calculations in quantum field theory. 
We show that trotterized time-evolution operators can be related by analytic continuation to the Euclidean transfer matrix on an anisotropic lattice. In turn, trotterization entails renormalization of the temporal and spatial lattice spacings. Based on the tools of Euclidean lattice field theory, we propose two schemes to determine Minkowski lattice spacings, using Euclidean data and thereby overcoming the demands on quantum resources for scale setting. 
In addition, we advocate using a fixed-anisotropy approach to the continuum to reduce both circuit depth and number of independent simulations. We demonstrate these methods with \texttt{qiskit} noiseless simulators for a $2+1$D discrete non-Abelian $D_4$ gauge theory with two spatial plaquettes.

\end{abstract}

\maketitle

\section{Introduction}

Inherent obstacles to classically simulating quantum field theories motivate developing quantum computer~\cite{Feynman:1981tf,Jordan:2017lea,pqa_loi}. For lattice-regulated theories, the exponential Hilbert space limits deterministic methods while stochastic methods grapple with sign problems.  These sign problems hamper calculations at finite-density~\cite{Gibbs:1986xg,Gibbs:1986ut,Troyer:2004ge,Alexandru:2018ddf} and in Minkowski spacetime~\cite{Alexandru:2016gsd, Hoshina:2020gdy}. While large-scale, fault-tolerant quantum computers will revolutionize our understanding of nature, for the foreseeable future, quantum computers will be limited to hundreds of non-error-corrected qubits with circuit depths less than 1000 gates-- the so-called Noisy Intermediate-Scale Quantum (NISQ) era.  Despite this, toy calculations in high energy physics ~\cite{Martinez:2016yna,Kokail:2018eiw,Klco:2018kyo,Lamm:2018siq,Macridin:2018gdw,Klco:2019evd,Gustafson:2019mpk,Gustafson:2019vsd,Kreshchuk:2020aiq} and nuclear physics~\cite{Dumitrescu:2018njn,Roggero:2018hrn,Lu:2018pjk} have been performed using existing quantum computers, representing the first step towards quantum simulating field theories.

Alongside the necessary hardware improvements, theoretical questions must be resolved to fully utilize a digital quantum computer.  Due to the finite resources, one must regulate the quantum field theory.
This regularization occurs in multiple steps: discretization, digitization, state preparation, propagation, and evaluation. Each can introduce new operators and potentially break symmetries. In addition, quantum noise can be interpreted as additional terms in the Hamiltonian.  In order to recover the physical theory, the resulting effects from regularization and quantum noises must be renormalized.

Following classical lattice field theory (LFT), it seems natural to first regularize the theory by discretizing spacetime. Then one could represent the (Minkowski or Euclidean) spacetime lattice in the qubits. This allows direct access to the entire path integral. The authors of~\cite{Temme:2009wa,Clemente:2020lpr} suggest this is useful for finite-density field theory. Alas, the number of qubits scales with the spacetime volume $V$ which improves the scaling of $e^V$ in classical computations. For time-dependent field theories, the preferred method is to use the Hamiltonian formalism. In this case, the number of qubits scales with a spatial lattice. Discretization reduces spacetime symmetries and introduce new operators into the LFT which are not present in the continuum theory that  modifies the nonperturbative renormalization.

For efficient digital simulations, truncation of the local lattice degrees of freedom is also necessary. Digitization represents the task of formulating, representing, and encoding QFTs for digital quantum computers.  Some natural encodings exist for fermionic degrees of freedom~\cite{Jordan:1928wi,Bravyi2002FermionicQC,Chen:2018nog}. Further proposals discuss how to map lattice fermions (e.g. Wilson and staggered) onto these encodings~\cite{Muschik:2016tws} or use gauge symmetry to eliminate the fermions~\cite{Zohar:2018cwb,Zohar:2019ygc}. The relative merits of each are only starting to be understood. The question of gauge boson digitization is murkier, with complicated tradeoffs~\cite{Hackett:2018cel,Alexandru:2019nsa,Singh:2019uwd,Singh:2019jog,Davoudi:2020yln,Bhattacharya:2020gpm,Barata:2020jtq,Kreshchuk:2020kcz,Ji:2020kjk}. Digitizing reduces symmetries -- either explicitly or through finite-truncations~\cite{Zohar:2013zla}. Care must be taken, as the regulated theory may not have the original theory as its continuum limit~\cite{Hasenfratz:2001iz,Caracciolo:2001jd,Hasenfratz:2000hd,PhysRevE.57.111,PhysRevE.94.022134,article}. A particularly illustrative example of the complications between truncations and renormalization can be found in~\cite{Wilson:1994fk}. Prominent proposals for digitization can be broadly classified~\cite{digi_loi} into: Casimir dynamics~\cite{Zohar:2012ay,Zohar:2012xf,Zohar:2013zla,Zohar:2014qma,Zohar:2015hwa,Zohar:2016iic,Klco:2019evd,Ciavarella:2021nmj} potentially with auxillary fields~\cite{Bender:2018rdp}, conformal truncation~\cite{Liu:2020eoa}, discrete groups~\cite{Hackett:2018cel,Alexandru:2019nsa,Yamamoto:2020eqi,Ji:2020kjk,Haase:2020kaj}, dual variables~\cite{PhysRevD.99.114507,Bazavov:2015kka,Zhang:2018ufj,Unmuth-Yockey:2018ugm,Unmuth-Yockey:2018xak}, light-front kinematics~\cite{Kreshchuk:2020dla,Kreshchuk:2020aiq} loop-string-hadron formulation~\cite{Raychowdhury:2018osk,Raychowdhury:2019iki,Davoudi:2020yln}, quantum link models~\cite{Wiese:2014rla,Luo:2019vmi,Brower:2020huh,Mathis:2020fuo}, and qubit regularization~\cite{Singh:2019jog,Singh:2019uwd,Buser:2020uzs}.

Given a digitization, the next obstacle is initializing strongly-coupled quantum states in terms of fundamental fields. Much of the literature emphasized ground state preparation~\cite{peruzzo2014variational,Kokail:2018eiw,Abrams:1998pd,nielsen2000quantum,PhysRevLett.117.010503,farhi2000quantum,Farhi472,Kaplan:2017ccd} but thermal and particle states have been investigated~\cite{2010PhRvL.105q0405B,Jordan:2011ne,Jordan:2011ci,Garcia-Alvarez:2014uda,Jordan:2014tma,Moosavian:2017tkv,Lamm:2018siq,Gustafson:2019mpk,Gustafson:2019vsd,Harmalkar:2020mpd,Gustafson:2020yfe,Jordan:2017lea,Klco:2019xro,PhysRevLett.108.080402,brandao2019finite,Clemente:2020lpr,motta2020determining,deJong:2021wsd}. For methods which construct states using regulated theories, careful study of the renormalization to properly match onto the physical limit is required~\cite{Maiani:1990ca,Bruno:2020kyl}.

Propagating for a time $t$ requires the unitary operator of $\mathcal U(t)=e^{-iHt}$ -- a generically dense matrix -- which cannot be efficiently constructed on a quantum computer. Instead, it must be approximated. A common method is trotterization, whereby $\mathcal U(t)\approx (e^{-iH' \frac{t}{N}})^N$ with an approximate Hamiltonian $H'$.  For some $H$, this allows for efficient simulations~\cite{Jordan:2011ne,Jordan:2011ci,Jordan:2017lea,Garcia-Alvarez:2014uda,Jordan:2014tma,Moosavian:2017tkv,Bender:2018rdp,haah2018quantum,Du:2020glq,PhysRevX.11.011020}. Most gauge theory studies consider the Kogut-Susskind Hamiltonian~\cite{PhysRevD.11.395}, but Hamiltonians with reduced lattice artifacts also exist~\cite{Luo:1998dx,Carlsson:2001wp} and deserve study.  Other approximations of $\mathcal U(t)$ exist: QDRIFT~\cite{PhysRevLett.123.070503}, variational approaches~\cite{cirstoiu2020variational,gibbs2021longtime,yao2020adaptive}, Taylor series~\cite{PhysRevLett.114.090502}, and qubitization~\cite{Low2019hamiltonian}. Initial resource comparisons have been performed for the Schwinger model~\cite{Shaw:2020udc}. Approximating $\mathcal U(t)$ can be understood as introducing $t$-translation violating operators into~$H'$.

There is little difficulty in evaluating the expectation values of instantaneous hermitian operators. Observables dependent on time-separated operators (e.g. parton distribution functions~\cite{Lamm:2019uyc,Kreshchuk:2020dla,Echevarria:2020wct}, particle decays~\cite{Ciavarella:2020vqm}, and viscosity~\cite{Cohen:2021imf}) are more complicated. Naively, the first measurement collapses the state, preventing further evolution.  Ways to overcome this include ancillary probe-and-control qubits~\cite{PhysRevLett.113.020505,Ortiz:2000gc,Lamm:2019bik,Lamm:2019uyc,Gustafson:2020yfe} and phase estimation~\cite{Abrams:1998pd,Roggero:2018hrn}. For time-separated matrix elements, it is yet unknown how to do
nonperturbative renormalization like RI/SMOM~\cite{Martinelli:1994ty,Aoki:2007xm,Sturm:2009kb} on quantum computers.

Noisy quantum devices can also be viewed as introducing new operators. The best-studied examples of this are related to gauge-violating operators~\cite{Stannigel:2013zka,Stryker:2018efp,Halimeh:2019svu,Lamm:2020jwv,Tran:2020azk,Halimeh:2020ecg,Halimeh:2020kyu,Halimeh:2020djb,Halimeh:2020xfd,VanDamme:2020rur,Kasper:2020owz,Halimeh:2021vzf}.  Which operators are introduced and which symmetries are broken are both hardware and digitization dependent.

In this paper we investigate
the renormalization of LFT in Minkoswki spacetime due to trotterizing $\mathcal{U}(t)$. The consequence of this will be shown to be the introduction of a temporal lattice spacing, and new operators depending upon it which vanish in the Hamiltonian limit.

In the continuum limit, Minkowski and Euclidean results are the analytic continuation of each other~\cite{Osterwalder:1973dx,Osterwalder:1974tc,Luscher:1984is}. At finite $a_t$ and finite statistics, this exact relation is complicated, but approximate relations remain~\cite{Giordano:2009ug,Giordano:2015fsa,Rossi:2018zkn,Briceno:2017cpo,Hoshina:2020gdy}. While analytic continuation of lattice observables suffer from signal-to-noise problems~\cite{Jarrell:1996rrw,Meyer:2011gj,Burnier:2013nla,Pawlowski:2017hpe,Briceno:2017cpo,Tripolt:2018xeo,Bulava:2019kbi,Karpie:2019eiq,Bruno:2020kyl}, observables suitable for scale setting have been studied~\cite{Asakawa:2000tr,Sasaki:2005ap}. 
Since knowledge of $a,a_t$ is required for any continuum extrapolation, performing scale setting with classical computations would significantly improve the common error budget of quantum computations as the uncertainties for scale setting could be reduced using Euclidean data. We will explore two different schemes for performing analytic continuation of the renormalized lattice spacings, and demonstrate its capabilities for reliable Minkowski scale setting through classical Euclidean computations. 

A crucial part of our study is to explore how Minokowski lattice observables computed with quantum circuits can be extrapolated to the continuum in an efficient manner. We will show that trotterized time-evolution can be understood as a Minkowski path integral on an anisotropic lattice\footnote{This point was first mentioned in~\cite{Kaplan:2018vnj}}. We present a toy model, a $D_4$ gauge group in $2+1$D with a two spatial plaquettes, to exemplify the power of a fixed anisotropy trajectory to extrapolate quantities to the continuum limit. This requires as a first step to establish the scale setting for the lattice spacings, $a,a_t$, that can profit from our analytic continuation schemes.

This paper is organized as follows. In Sec.~\ref{sec:elft} we briefly review the Euclidean action lattice formalism and its connection through the transfer matrix to the Hamiltonian formalism. In Sec.~\ref{sec:mlft} we derive the trotterized real-time evolution operator and relate it to the transfer matrix. Based on this, we propose two schemes to obtain Minkowski lattice spacings via analytic continuation and advocate the use of a fixed-anisotropy approach to the continuum.
In Sec.~\ref{sec:errorAC}, we discuss the 
systematic errors from computing $a,a_t$ via analytic continuation. Further, in Sec.~\ref{sec:num} we present numerical results in our toy model for these techniques. Finally, we conclude in Sec.~\ref{sec:concl}.

\section{Lattice basics}
\label{sec:elft}
To understand how renormalization arises in quantum simulations, it is useful to review the connection between the Kogut-Susskind Hamiltonian~\cite{PhysRevD.11.395} and the Euclidean Wilson action.  We summarize the derivation of~\cite{Creutz:1984mg} that begins with the anisotropic Wilson action in Euclidean time $\tau=it$ defined on a spacetime lattice:
\begin{equation}\label{eq:Wilson}
    S_E=-\beta_t\sum_{t}\re\Tr U_t-\beta_s\sum_{s}\re\Tr U_s
\end{equation}
where $i= t,s$ refers to temporal and spatial plaquettes $U_i$ formed from gauge links given by elements of the group.  The anisotropy comes from using different couplings for spatial and temporal plaquettes, that can be written as
\begin{align}
\label{eq:betas}
    \beta_t(a,a_0)=\frac{a}{g^2_t(a,a_0) a_0},\qquad    \beta_s(a,a_0)=\frac{a_0}{g^2_s(a,a_0) a}
\end{align}
with $\beta_i(a,a_0),g_i(a,a_0)$  depending nonperturbatively on the temporal and spatial lattice spacings, $a_0,a$.

The first step in the process of computing physical observables from LFT is the determination of the lattice spacings through \textit{scale setting}. For simplicity we will consider the isotropic case via $\beta_E\equiv\beta_t=\beta_s$, $a=a_0$ and in analogy with Eq.~(\ref{eq:betas}), define $\beta_E=2N g_E^{-2}$ for $SU(N)$ group. The anisotropic case merely requires performing the procedure for both $a,a_0$ independently. One scale sets by computing a lattice quantity $am(\beta_E)$ where $m$ has a known physical value $m^{\rm phys}$ (e.g. the pion mass). Any lattice $m(\beta_E)$ differs from the true $m^{\rm phys}$ by $a$ dependent errors, but for this one specific observable we set $m(\beta_E)=m^{\rm phys}$ to obtain a dimensionful value for $a$
\begin{equation}
    a = \frac{[a m(\beta_E)]}{m^{\rm phys}}
\end{equation}
With this, $\beta_E$ is removed from our theory and we can speak only in terms of $a$.  All other lattice masses can then be written as $a m_k(a)$ and their continuum values can be predicted by computing them at multiple lattice spacings and extrapolating to $a\rightarrow 0$ via
\begin{equation}
    \frac{m_k^{\rm phys}}{m^{\rm phys}}=\frac{a m_k(a)}{am(a)}+O(a^n)
\end{equation}

When working with continuous gauge theories, there is no theoretical issue with computing at arbitrarily small $a$, but one is limited by computing resources due to topological freezing and critical slowing-down. On quantum devices, the current resources require dramatic approximations of the continuous group.  Here, we will consider the discrete gauge theories. Certain discrete subgroups of continuous groups are effective field theories for the continuous groups~\cite{Fradkin:1978dv,Horn:1979fy} which break down below a minimum lattice spacing~\cite{Hackett:2018cel,Alexandru:2019nsa,Ji:2020kjk}. Therefore one cannot take $a$ of a discrete group arbitrarily close to zero.

Lattice quantities like $am(\beta_E)$ are obtained from correlation functions, e.g. the temporal correlator $\langle\mathcal{O}_i(n a_0) \mathcal{O}_j(0)\rangle$.  In the limit where the temporal length of the lattice goes to infinity, this correlator becomes a vacuum expectation value 
\begin{align}\label{eq:space correlator}
    \langle\mathcal{O}_i(n a_0)\mathcal{O}_j(0)\rangle&=\sum_k \langle 0| \mathcal{O}_i|k\rangle\langle k |O_j|0\rangle e^{-n a_0 m_k}
    \end{align}
From these correlators, one extracts $a_0 m_k$ which correspond to the lattice eigenenergies. For scale setting, one usually wants the lowest energy state $a_0 m_1$ of a specific sector, which can be extracted from the sum by taking $n$ large:
\begin{align}
\label{eq:euccor}
    \langle\mathcal{O}_i&(n a_0)\mathcal{O}_j(0)\rangle\notag\\&=\langle 0|\mathcal{O}_i|1\rangle\langle 1 |O_j|0\rangle e^{-n a_0 m_1}+O(e^{-na_0 \Delta E}).
\end{align}
with $\Delta E$ the energy gap between the lowest energy state and the next lowest energy state.

An equivalent way of expressing the renormalized parameters is by defining the anisotropic parameter $\xi \equiv a/a_0$ and considering $\xi$ and $a$ as independent parameters. By allowing $\xi\neq 1$ and especially $\xi \gg 1$, lattice practitioners have achieved great success with probing glueballs~\cite{Morningstar:1996ix, Byrnes:2003gg}, high temperature thermodynamics~\cite{Bralic:1989du}, etc. As we approach the Hamiltonian limit ($a_0\rightarrow0)$, another couplings, $g_H^2=g_s g_t$, and the speed of light, $c=g_s g_t^{-1}$, become useful. These bare couplings are related to each other in the weak coupling limit by~\cite{Hasenfratz:1981tw,Karsch:1982ve}
\begin{align}\label{eq:renorm}
    g^{-2}_s(a,a_0)&=g_E^{-2}(a)+c_s(\xi)+O(g^2_E)\notag\\
     g^{-2}_t(a,a_0)&=g_E^{-2}(a)+c_t(\xi)+O(g^2_E)\notag\\
     g^{-2}_H(a,a_0)&=g_E^{-2}(a)+\frac{c_t(\xi)+c_s(\xi)}{2}+O(g^2_E)\notag\\
    c(a,a_0)=&1+\frac{c_t(\xi)-c_s(\xi)}{2}g_E^2(a)+O(g_E^4)
\end{align}

The $c_i(\xi)$ were computed perturbatively for $SU(N)$ at $\xi=\infty$ for $D=4$ in \cite{Hasenfratz:1981tw}.  This was generalized in \cite{Karsch:1982ve} to arbitrary $\xi$ and to arbitrary dimensions in \cite{Hamer:1996ub}.  

For typical values of $\beta_i$ considered in simulations, there are large corrections to these weak coupling results and thus nonperturbative determination of $a,\xi$ is required~\cite{Morningstar:1996ze,Morningstar:1997ff,Alford:2000an}. In pure gauge theory, one method for the determination of $\xi$ is made by comparing ratios of spatial-spatial Wilson loops to spatial-temporal Wilson loops~\cite{Klassen:1998ua}. Once $\xi$ is measured, $a$ could be determined using standard methods such as the Sommer scale $r_0$~\cite{Morningstar:1996ix,Byrnes:2003gg,Chen:2005mg} or the Wilson flow~\cite{Luscher:2010iy,Borsanyi:2012zr}. 

Euclidean lattice theories satisfying the reflection positivity have a well-defined Hamiltonian with real eigenvalues~\cite{Osterwalder:1973dx,Luscher:1976ms}. To the derive this Hamiltonian, we first define a transfer matrix, $T(a,a_0)$ which takes a state at time $\tau$, $|\tau\rangle$, to $|\tau+1\rangle$.  $T$ is related to the action through the partition function $Z$:
\begin{align}
    Z=\int D Ue^{-S_E}=\tr T(a_0)^N
\end{align}
where $N$ is the number of temporal lattice sites. It follows that the matrix elements of $T(a_0)$ are~\cite{Creutz:1984mg}
\begin{align}\label{eq:transfer}
    \langle \tau+&1|T(a_0)|\tau\rangle\nonumber\\=&e^{\frac{\beta_s}{2}\sum_{s}\re \tr  U_s}e^{\beta_t\sum_{\{\tau,\tau+1\}} \re \tr U_t}e^{\frac{\beta_s}{2}\sum_{s}\re \tr  U_s}\notag\\
    \equiv&T_{V}^{1/2}T_KT_{V}^{1/2},
\end{align}
where we have symmetrically split the potential term. In order to extract a Hamiltonian from $T(a_0)$, it is  convenient to have $T(a_0)$ only in terms of a single time slice. While for $U_s$ this presents no issues, $U_t$ couples the same link at two times $U_{ij}(\textbf{x},\tau),U_{ij}(\textbf{x},\tau+1)$ via two time-like links $U_{0i}(\textbf{x},\tau),U_{0j}(\textbf{x}+1,\tau)$. Fixing into the temporal gauge, $U_{0i}=\mathbb{1}$, yields for the kinetic term in the action
\begin{equation}
    S_{K}=-\beta_t\sum_{\{\tau,\tau+1\}}\re\Tr U_{ij}(\tau)U_{ij}^\dag(\tau+1)
\end{equation}
To proceed, we need to remove the dependence on $U_{ij}^\dag(\tau+1)$ and express $T(a_0)$ in terms of operators.  The link operator is easy to define $\hat{U}_{ij}|\tau\rangle={U}_{ij}|\tau\rangle$.  Therefore 
\begin{equation}
T^{1/2}_V=e^{\frac{\beta_s}{2}\sum_s\re\tr \hat{U}_s}.
\end{equation}
For $T_K$, we need an operator that changes a given link, 
\begin{equation}
    R_{ij}(g)|\tau\rangle =|\tau'\rangle,\quad\text{where } U_{ij}\rightarrow g U_{ij}
\end{equation}
this operator has the group property of $R_{ij}(g)R_{ij}(h)=R_{ij}(gh)$.  This gauge link translation can be used to define a conjugate momentum to $\hat{U}_{ij}$ by performing a rotation on $U_{ij}(\textbf{x},\tau+1)$. With this, we write 
\begin{equation}
T_K=\prod_{\{ij\}}T_{K,ij}=\prod_{\{ij\}}\left[\int Dg R_{ij}(g) e^{\beta_t\re\tr g}\right],
\end{equation}
where the product is over all spatial links $U_{ij}(\tau)$.  Any group element can be written as $g=e^{i\omega\cdot\lambda}$ where $\lambda_i$ are the adjoint generators, and $R_{ij}(g)=e^{i\omega\cdot l_{ij}}$ can be written in terms of the generators $l_{ij}$ for that representation.  Defining $\prod_\alpha(D\omega^\alpha)J(\omega)$ as the invariant group measure with a Jacobian $J$, it is possible to rewrite $T_K(a_0)$ as
\begin{align}
\label{eq:tkijaprox}
    T_{K,ij}=\int \prod_\alpha(D\omega^\alpha)J(\omega) e^{i\omega\cdot l_{ij}} e^{\beta_t\tr \cos(\omega\cdot \lambda)}
\end{align}
This transfer matrix is exact for any $a_0$.  At the cost of a sum over all character functions of the group, it can be performed analytically. This is done in~\cite{Lamm:2019bik} and appears practical for discrete groups. 
But for continuous group, the summation is over infinite terms which is undesirable.

As $a_0\rightarrow 0$ for unitary continuous group, Eq.~(\ref{eq:tkijaprox}) can be expanded to $O(\omega^2)$:
\begin{align}
    &T_{K,ij}=\int \prod_\alpha(D\omega^\alpha)\left(1+O(\omega^2)\right)e^{i\omega\cdot l_{ij}+\beta_t[\tr \mathbb{1}-\frac{1}{4}\omega^2+O(\omega^4)]}
\end{align}
leaving Gaussian integrals. Integrating yields 
\begin{eqnarray}\label{eq:Tk0}
    T_K&=&\mathcal{N}e^{-\beta_t^{-1}\sum_{\{ij\}}l_{ij}^2}
\end{eqnarray}
where $\mathcal{N}$ is an overall normalization. For discrete groups, the contribution to $T_K$ is also dominated by group elements close to $\mathbb{1}$ when $a_0\to 0$. However, since for discrete group $g$ cannot be arbitrarily close to $\mathbb{1}$, a naive limit of taking $a_0\to 0$ leads to degenerate spectrum for $T_K$. Special care has to be taken~\cite{Harlow:2018tng} to avoid this degeneracy as we also show for the $D_N$ group in Appendix~\ref{sec:discrete}. 

\begin{figure*}[t]
\includegraphics[width=\linewidth]{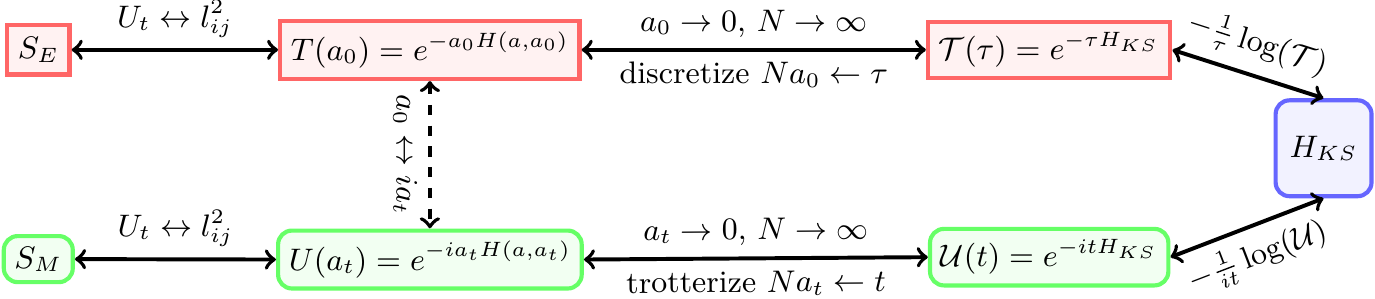}
\caption{Schematic of the relations between the various lattice and continuum functions discussed in this work.}
\label{fig:relations}
\end{figure*}

Neglecting the normalization factor $\mathcal{N}$, the final transfer matrix $T(a_0)$ is given by
\begin{equation}\label{eq:Ta0}
    T(a_0)=e^{\frac{\beta_s}{2}\sum_s\re\tr \hat{U}_s}e^{-\beta_t^{-1}\sum_{\{ij\}}l_{ij}^2}e^{\frac{\beta_s}{2}\sum_s\re\tr \hat{U}_s}.
\end{equation}
Since $T(a_0)$ corresponds to the translation from $\tau$ to $\tau+1$, it can be used to define a Hamiltonian $H(a_0, a)$ These steps form the link between $S_E$ and $T(a_0)$ in Fig.~\ref{fig:relations}.
\begin{align}
\label{eq:tfin}
    T(a_0)\equiv e^{-a_0 H(a,a_0)}.  
\end{align}
However, because $l_{ij}$ and $\hat{U}_{ij}$ are non-commuting operators, writing Eq.~(\ref{eq:Ta0}) as a single exponential requires application of the Baker-Campbell-Hausdorff (BCH) formula:
\begin{align}
\label{eq:bch}
   e^{t X}e^{tY}e^{tX} = e^{t(2X + Y) -\frac{t^2}{6}([X,[X,Y]]-[Y,[X,Y]])+\hdots}.
\end{align}
Using this, we obtain for $H(a,a_0)$:
\begin{widetext}
\begin{align}
\label{eq:latham}
    H(a,a_0)=&\frac{1}{c(a,a_0)a}\bigg(g_H^2(a,a_0)\sum_{\{ij\}}l_{ij}^2-g^{-2}_H(a,a_0)\sum_s\re\tr \hat{U}_s\notag\\
    &-\frac{1}{24}\frac{1}{c^2(a,a_0)\xi^2}\sum_{\{ij\},s}\left(g_H^{2}(a,a_0)[2l_{ij}^2,[l_{ij}^2,\re\tr \hat{U}_s]]-g_H^{-2}(a,a_0)[\re\tr\hat{U}_s,[l_{ij}^2,\re\tr \hat{U}_s]]\right)+\hdots\bigg)
\end{align}

\end{widetext}
For conciseness, we define certain Hamiltonian terms
\begin{align}
\label{eq:hdef1}
    H_K(a,a_0)&=\frac{g_H^2(a,a_0)}{c(a,a_0)a}\sum_{\{ij\}}l_{ij}^2\notag\\
    H_V(a,a_0)&=-\frac{1}{g_H^2(a,a_0)c(a,a_0)a}\sum_s\re\tr \hat{U}_s
    \end{align}

It is important to emphasize that varying either $a_0$ or $a$ requires an adjustment to $c(a,a_0)$ and $g_H(a,a_0)$ to preserve the scale setting condition. Taking the continuous-time limit of the transfer matrix: 
\begin{equation}
    \mathcal{T}(\tau)\equiv \lim_{a_0\rightarrow 0, N\rightarrow\infty}T(a_0)^N,
\end{equation}
the BCH terms in $H(a,a_0)$ vanish and we obtain the Kogut-Susskind Hamiltonian~\cite{PhysRevD.11.395}, $H_{KS}\equiv-\frac{1}{\tau}\log(\mathcal{T}(\tau))$ (See Fig.~\ref{fig:relations}):
\begin{equation}
\label{eq:ksham}
    H_{KS}=\frac{1}{c(a)a}\left(g_H^2(a)\sum_{\{ij\}}l_{ij}^2-\frac{1}{g_H^2(a)}\sum_{s}\re\Tr U_s\right).
\end{equation}
Besides the $\xi$-dependent terms, another difference between $H_{KS}$ and $H(a,a_0)$ is that $c,g_H$ only depend upon $a$ in Eq.~(\ref{eq:ksham}), while in Eq.~(\ref{eq:latham}) they also depend on $a_0$.

Historically, the Hamiltonian formalism with limited success was used to evaluate lattice theory by computing results analytically~\cite{Ligterink:2000ug,McIntosh:2001fm,Wichmann:2001hv,Jiang:2012zzk}, variationally~\cite{Beccaria:2000eg,Carlsson:2002ss,Carlsson:2003wx,Carlsson:2003fb}, and with exact diagonalization~\cite{Hiller:2016itl}. In such cases, there was no benefit to keeping $a_0$ finite and therefore all were done in the Hamiltonian limit. In contrast, quantum simulations have good reasons to considering the Hamiltonian at finite $a_0$ as we demonstrate in the next section. 

\section{Trotterization and Time-Evolution}
\label{sec:mlft}
The starting point for real-time evolution on quantum computers is to define $\mathcal{U}(t)=e^{-iHt}$. For gauge theories, one typically takes the lattice Hamiltonian to be $H_{KS}$ in Eq.~(\ref{eq:ksham}). This $\mathcal{U}(t)$ cannot be implemented easily on quantum computers and must be approximated, resulting in effects that have to be renormalized. Trotterization discretizes the evolution into $N=t/a_t$ steps formed of products of $e^{i x_i H_j}$ where different choices of $x_i\propto a_t$ lead to errors at different order of $a_t$ and $H_{i}$ are a decomposition of $H$ into mutually noncommuting terms. In the case of $H_{KS}$, there are two $H_i=H_K,H_V$, in which case $\mathcal{U}(t)$ is approximated by
\begin{align}
    \mathcal{U}(t)&=e^{-iH_{KS}t}\notag\\&\approx [e^{ix_1 H_{V}}e^{ix_2 H_{K}}e^{ix_3 H_{V}}e^{ix_4 H_{K}}\cdots]^N+O(a_t^{p}).
\end{align}
The number of terms and $x_i$ for a given $O(a_t^p)$ error can be derived by repeated applications of the BCH relation, Eq.~(\ref{eq:bch}). In the case of $O(a_t^2)$, this corresponds to
 \begin{equation}
 \label{eq:utrot}
        U(a_t) \approx e^{-i a_t H_V/2}e^{-i a_t H_K}e^{-i a_t H_V/2}
 \end{equation}
 Following Fig.~\ref{fig:relations}, we see that we have essentially reversed the path taken to derive $\mathcal{H}_{KS}$ from $T(a_0)$. With this approximation, we can ask, what Hamiltonian are we actually evolving with respect to? Analogous to the euclidean derivation of Eq.~(\ref{eq:latham}) from Eq.~(\ref{eq:Ta0}), we can define
 \begin{equation}
 \label{eq:ufin}
 U(a_t)=e^{-i a_t H(a_t)},
 \end{equation}
 finding
 \begin{align}
 \label{eq:timham}
     &H(a_t) =H_K + H_V \notag\\
    &-\frac{a_t^2}{24}\bigg([2H_K,[H_K, H_V]] + [H_V,[H_K,H_V]] \bigg)+\hdots
 \end{align}
 
 By trotterizating $\mathcal{U}(t)$, we have introduced a temporal lattice spacing $a_t$. One might be tempted to believe that the renormalized $a_t$ is a parameter that can be directly tuned, but this is incorrect.  This can be seen by inserting \eq{hdef1} into \eq{timham}, from which one observes that in the same way as the Euclidean results, $a_t$ is always multiplied by $[c(a,a_t) a]^{-1}$.  Thus, changes in $a_t$ are compensated by modifying the bare speed of light.  It it therefore natural to define a lattice bare parameter which we can control in simulations:
 \begin{equation}
 \label{eq:dt}
     \delta_t\equiv\frac{a_t}{c(a,a_t)a}.
 \end{equation}
Thus we find that $a_t$ must be determined nonperturbatively by scale setting. 
Naively, $a_t,a$ would be obtained by performing a quantum simulation, which on near-term hardware is likely to be noisy, and therefore the lattice spacings will have large uncertainties.  Since all other lattice observables depend upon the scale-setting, minimizing the uncertainties of $a,a_t$ is crucial to the overall program of quantum simulations of LFT.  In practice, keeping these uncertainties small require high statistics of very deep circuits with error mitigation. Instead of computing $a_t,a$ on the quantum device, it is possible to utilize classical Euclidean computations of $a,a_0$ to scale setting in a quantum computer.

If the Minkowski and Euclidean lattice Hamiltonians were the same, we could trivially set $a,a_0$ to $a$ and $a_t$ -- establishing a link between the lattice results in each metric (See Fig.~\ref{fig:relations}). However, the Hamiltonians only match when $a_t,a_0\rightarrow0$, when they both reduce to $H_{KS}$. Instead, at finite $a_t,a_0$ they differ as we show explicitly in the following. 
In analogy to \eq{dt}, we define the quantity $\delta_\tau\equiv \frac{a_0}{c(a,a_0)a}$ on the Euclidean side. With this, we can recast $T(a_0)$ in \eq{Ta0}:
\begin{eqnarray}
\label{eq:tranferbare}
   T(\delta_\tau, g^2_H)&=& e^{-\delta_\tau \oh_V/2}e^{-\delta_\tau \oh_K }e^{-\delta_\tau \oh_V/2}
\end{eqnarray}
with the following Hamiltonian terms:
    \begin{align}
    \label{eq:hdef2}
    \oh_K&=g_H^2\sum_{\{ij\}}l_{ij}^2,\quad\oh_V=-g^{-2}_H\sum_s\re\tr \hat{U}_s.
\end{align}
Taking $\delta_\tau\rightarrow i\delta_t$ in \eq{tranferbare}, we analytically continue $T(\delta_\tau,g_H^2)$, obtaining
\begin{eqnarray}\label{eq:U}
   U(\delta_t, g^2_H) &&=e^{-i \delta_t \oh_V/2}e^{-i\delta_t \oh_K}e^{-i \delta_t \oh_V/2}
\end{eqnarray}
which we recognize as \eq{utrot} written in terms of bare parameters. From these, we can define the dimensionless Hamiltonians using only bare parameters:
\begin{align}
\label{eq:HIlatt}
&\oh(\delta,g_H^2) =\oh_K + \oh_V \notag\\
    +&\frac{\delta^2}{24}\bigg([2\oh_K,[\oh_K, \oh_V]] + [\oh_V,[\oh_K,\oh_V]]\bigg)+\hdots
\end{align}
where $\delta=\delta_\tau, i\delta_t$ depending on the metric signature. From this we see that there are differing signs for the BCH terms in real and imaginary time.  Given that the Hamiltonians differ, correlation functions and the scale setting observables $a_t m$ and $a_0 m$ must also differ even if we take $\delta_t=\delta_\tau$. But, these differences arise at $O(\delta_t^2,\delta_\tau^2)$ and vanish in the continuous time limit, $\delta_t=\delta_\tau=0$. 

One possible scheme for using $a,a_0$ to determine $a,a_t$ would be to simply neglect these $O(\delta^2_t,\delta^2_\tau)$ errors and assume the two sets of scales are equal for the same $g_H$ and $\delta_\tau=\delta_t$:
\begin{align}
     \mathbf{Scheme~A}:\quad&
     a(\delta_\tau, g_H^2)\rightarrow a(\delta_t, g_H^2)\notag\\
   & a_0(\delta_\tau, g_H^2)\rightarrow a_t(\delta_t, g_H^2)
\end{align}
A benefit of this scheme is that only one set of Euclidean couplings is simulated. While this scheme introduces an $O(\delta_t^2,\delta_\tau^2)$ systematic error into the scale setting, one could easily imagine it being tolerable compared to errors from near-term quantum computers.

In principle, this systematic error can be reduced by observing that if one takes $\delta_\tau \rightarrow i\delta_t$, then the two Hamiltonians agree. Formally, this means that the eigenvalues $m_k(\delta_t, g_H^2)$ are the analytic continuation of $m_k(\delta_\tau, g_H^2)$.

While the spatial correlators in Minkowski behave exactly like the Euclidean ones of \eq{space correlator} with the replacement of $m_k(\delta_t, g_H^2)$, the temporal correlators require $a_0\rightarrow i a_t$
\begin{align}\label{eq:mincor}
    \langle\mathcal{O}_i(n a_t)\mathcal{O}_j(0)\rangle&=\sum_k \langle 0| \mathcal{O}_i|k\rangle\langle k |O_j|0\rangle e^{-i n a_t m_k}
    \end{align}
where $|k\rangle$ are the Minkowski eigenstates.  While the excited state contributions do not decrease with $na_t\rightarrow \infty$, provided that we can isolate a single scale setting parameter $m$ then the lattice results $a_t m(a,a_t)$ are the analytic continuation of $a_0 m(a,a_0)$.  This suggest that a scale setting scheme with reduced systematic error is through analytical continuation:
\begin{align}
  \mathbf{Scheme~B}:\quad&a(\delta_\tau\rightarrow i\delta_t, g_H^2)\rightarrow a(\delta_t, g_H^2)\notag\\
   & a_0(\delta_\tau\rightarrow i\delta_t, g_H^2)\rightarrow a_t(\delta_t, g_H^2)
\end{align}

In contrast to Scheme A, this procedure requires the determination of the lattice spacings at multiple values of $g_H,\delta_\tau$ in the region around the desired lattice spacings.  With this set of values, one derives a fit function for $a(\delta_\tau, g_H^2), a_0(\delta_\tau, g_H^2)$.  This function can then be analytically continued to Minkowski space, reducing the nonperturbative BCH errors. The effectiveness of this method, like all analytic continuations of lattice results, depends sensitively on the statistics and fitting function, as we will discuss later.

In the preceding discussion, the relation between the lattice spacings depending upon the Hamiltonians being analytical continuations of each other.  Traditionally, Euclidean calculations are performed with an action like \eq{Wilson}.  Different actions correspond to different lattice Hamiltonians.  For the Wilson action, we observe that $H(a,a_0)$ of \eq{latham} is not the exact lattice Hamiltonian being computed, but arises only when we expand $T_K$ to $O(\omega^2)$.  Thus, a systematic mismatch occurs if one tries to scale set $a,a_t$ with Wilson action results. Furthermore, including the higher order $\omega$ terms from \eq{Wilson} leads to a non-trivial dependence on $\delta_\tau$. This causes a non-Hermitian Hamiltonian upon analytic continuation~\cite{Kanwar:2021tkd} although this behavior may prove manageable~\cite{Luscher:1984is}.  In contrast, using an action with a heat-kernel kinetic term (the Laplace-Beltrami operator)~\cite{Menotti:1981ry} with a Wilson single plaquette potential term, the higher order $\omega$ terms vanish and the mapping is exact.  Another approach to obtain a Hermitian lattice Hamiltonian useful in Minkowski space is obtained by analytic continuing the character expansion of $T_K (\delta_\tau=\delta_t, g^2_H)$ term-by-term~\cite{Hoshina:2020gdy,Kanwar:2021tkd}. 

With observables computed at multiple $a_t,a$, one can perform continuum extrapolations analogous to the Euclidean calculations. Influenced by nonrelativistic results, the literature has emphasized first approaching the Hamiltonian limit $a_t\rightarrow 0$, then taking $a\rightarrow 0$. This procedure introduces two inefficiencies. First, taking a continuum limit as a two-step procedure requires $m\times n$ separate simulations at $m$ values of $a_t$ for $n$ values of $a$ in the $(a,a_t)$ plane, extrapolating each fixed-$a$ set to the $a_t\rightarrow 0$, and then extrapolating the remaining $a$-dependent results to $a\rightarrow 0$. Secondly, because uncertainty in the $a\rightarrow 0$ extrapolation is controlled by how well each $a_t\rightarrow 0$ extrapolation is, one desires smaller $a_t$. This increases gate costs $\propto \delta_t^{-1}$.

Instead of first approaching the Hamiltonian limit, a more efficient trajectory is to compute a set of points $a,a_t\rightarrow 0$ at fixed $\xi_t$. This clearly reduces the total number of quantum simulations required. Additionally, since only one extrapolation is performed lattice errors are easier to control. This would allow larger $a_t$.  We thus expect the fixed-$\xi_t$ trajectory to avoid deep quantum circuits and suffer less from noise.

\section{The errors of scale-setting in Minkowski metric}
\label{sec:errorAC}
Although the analytic continuation between Euclidean and Minkowski scales proposed in \textbf{Scheme B} is formally exact, in practice one can only perform the analytic continuation from fits to discrete, noisy Euclidean data $a(\delta_\tau, g_H^2),a_0(\delta_\tau, g_H^2)$. This leads to a signal-to-noise problem when performing the analytic continuation~\cite{Pawlowski:2017hpe}. Indeed, to achieve a certain precision the Euclidean data has to be exponentially more accurate: intuitively, this is because excitations caused by BCH operators decay exponentially in $\tau$ but oscillate in $t$. As a result, low-energy observables in Euclidean lattices tend to be less sensitive to the variation of $\delta$ than their Minkowski counterparts.  
Hence the analytic continuation is ill-posed~\cite{Miller:1970SIAM}. Fortunately, at small $\delta$, the calculations with lower-energy states are less influenced by the higher order BCH operators, and hence one can reasonably tame the errors intrinsic to the Euclidean data. This observation is of crucial relevance for our scale-setting procedure.

On the other hand, the difference between Minkowski and Euclidean renormalized lattice scales must be smaller for smaller trotter steps $\delta_{t,\tau}$ as both $\oh(\delta_\tau, g_H^2)$ and $\oh(i\delta_t, g_H^2)$ are closer to the same Hamiltonian limit. This implies that there might be a parameter space where \textbf{Scheme A} yields more accurate results for Minkowski scale-setting, leading to question the feasibility of \textbf{Scheme B}.
In this section, we give upper bounds for the scale-setting error of both schemes and discuss if the advantage of analytic continuation of a given scheme is balanced out by its errors.

Consider the bare eigenvalue $\lambda(\delta_\tau,g_H^2)\equiv a_0( \delta_\tau,g_H^2)m^{\text{phys}}/\delta_\tau$ of $\oh(\delta_\tau, g_H^2)$. In real time, the equivalent eigenvalue of $\oh(i\delta_t, g_H^2)$ is $\lambda ( i\delta_t,g_H^2)\equiv a_t( \delta_t,g_H^2)m^{\text{phys}}/\delta_t$. 
Let $\lambda_m ( \delta_\tau,g_H^2)$ be the measured  values of $\lambda ( \delta_\tau,g_H^2)$ on Euclidean lattices, with a deviation from the theoretical value $\epsilon_A =|\lambda_m ( \delta_\tau,g_H^2)-\lambda (\delta_\tau,g_H^2)|$ from statistical errors of the measurement.

For $\delta_\tau = \delta_t$, \textbf{Scheme A} approximates $\lambda ( i\delta_t,g_H^2)$ as $\lambda_m ( \delta_\tau,g_H^2)$ such that the error is given by

\begin{align}
\label{eq:schemea}
   |\lambda&( i\delta_t,g_H^2)-\lambda_m ( \delta_\tau,g_H^2)|
   \notag\\
     &\leq  |\lambda( i\delta_t,g_H^2)-\lambda ( 0,g_H^2)| +|\lambda( \delta_\tau,g_H^2)-\lambda ( 0,g_H^2)|\notag\\
    &\quad+|\lambda( \delta_\tau,g_H^2)- \lambda_m ( \delta_\tau,g_H^2)|
\end{align}
where $\lambda (0, g_H^2)$ is the corresponding energy gap evaluated in the continuous time limit $\delta=0$. The first two terms on the RHS of \eq{schemea} quantify the errors from the BCH contributions.
The last term is $\epsilon_A$, the statistical error of the Euclidean temporal scale at $\delta_\tau$. 
We obtain the following constraint on the trotterization error according to the Bauer-Fike theorem \cite{Bauer-Fike1960}, 
\begin{align}
\label{eq:trottererror}
    |\lambda(\delta,g_H^2)-\lambda ( 0,g_H^2)|&\leq 2 \| \oh( \delta,g_H^2)-\oh(0,g_H^2)\|
\end{align}

At small $\delta$, $ \| \oh( \delta,g_H^2)-\oh(0,g_H^2)\|$ is dominated by the leading order BCH commutators of order $|\delta|^2$. 
\begin{widetext}
\begin{align}\label{eq:commutator norm}
    \|\oh(\delta, g_H^2)-\oh(0, g_H^2)\| &\lesssim  \frac{|\delta|^2}{24}\left (2\| [[\oh_V, \oh_K],\oh_K]\|+\|[[\oh_V,\oh_K],\oh_V]\|\right)\notag\\
    & \leq \frac{|\delta|^2}{3}\mathcal{N}_{link} (d-1) d_U \|\hat{l}^2\|\left (g_H^{2}8\|\hat{l}^2\|+ g_H^{-2}2(d-1)d_U\right) \leq \frac{|\delta|^2}{4(\max \delta)^2} M
\end{align}
\end{widetext}
where $d$ is the number of spacial dimensions, $d_U$ is the dimension of the representation of the group element $U$ in $H_V$, $\mathcal{N}_{link}$ is the number of links in the spacial lattice, and $\| \hat{l}^2\|$ is the spectral norm of $\hat{l}^2$. We have introduced additional definitions into the second line,
\begin{align}\label{eq:delta_max}
    \max \delta &\equiv\min\left\{\frac{g_H^2}{4(d-1)d_U}, \frac{1}{g_H^28\|\hat{l^2}\|}\right\}\\
    M&\equiv 2\mathcal{N}_{link}(d-1)d_U\|\hat{l}^2\|\max\delta 
\end{align}
The inequality of \eq{commutator norm} is only guaranteed within the range $|\delta|<\max \delta$ because next to leading order commutators are only negligible within such range, as shown in Appendix~\ref{ap:perturbativity}. 
Thus, \eq{trottererror} is bounded by $M/2$ for $|\delta |\leq \max \delta$. Combining all the above definitions and inequalities, the upper bound of temporal scale-setting errors for $\textbf{Scheme A}$ is
\begin{align}\label{eq:SchemeA}
    |\lambda(i\delta_t, g_H^2)&-\lambda_m (\delta_\tau, g_H^2)|\lesssim
   \bigg(\frac{\delta_t^2}{ \max\delta^2}  + \frac{\epsilon_A}{M}\bigg)  M
\end{align}
which is bounded by $M + \epsilon_A$ for $\delta \leq \max\delta$. As $M$ quantifies the upper limit of the trotterization error in \eq{trottererror} and also the error of \textbf{Scheme A}, we refer to $M$ as the error bound parameter from the BCH expansion.

As \textbf{Scheme B} requires the knowledge of the functional dependence of the scales on $g_H$ and $\delta_\tau$, we assume that the theoretical $\lambda(\delta, g_H^2)$ is both analytic and even in power of $\delta$ within the radius $|\delta|\leq \max \delta$, i.e. $\lambda(\delta,g_H^2)=\lambda(-\delta,g_H^2)$. This is based on the following perturbative argument: for small $|\delta|$, the BCH commutators $ \oh(\delta, g_H^2)-\oh(0, g_H^2)$ can be treated as perturbations to the Hamiltonian limit $\oh(0,g_H^2)$. With the second-order trotterization, all the non-vanishing BCH commutators depend on even orders of $\delta$. Therefore, perturbatively, the difference in the spectra of $\oh(\delta, g_H^2)$ and $\oh(0, g_H^2)$ should be analytic and even in powers of $\delta$ order by order. One thus fits the functional form $ \lambda_f(\delta_\tau, g_H^2)$ for Euclidean temporal scales in even powers of $\delta_\tau$. This guarantees that the analytic continuation $ \lambda_f(i\delta_t, g_H^2)$ is real, thus avoiding nonunitarity in the Minkowski metric.
 
Define $\epsilon_B$ as the maximum deviation of $ \lambda_f(\delta_\tau, g_H^2)$ from the theoretical $\lambda(\delta_\tau, g_H^2)$ across the Euclidean region, such that $|\lambda_f(\delta_\tau, g_H^2) - \lambda(\delta_\tau, g_H^2)|\leq \epsilon_B$ for all  $0<\delta_\tau\leq \max\delta$. $\epsilon_B$ is affected by both the precision of the measurements on the Euclidean data and the fitting procedure. Since the computational resources required grow with decreasing $\delta_\tau$, $|\lambda_f(0, g_H^2) - \lambda(0, g_H^2)|$ is likely to be the largest deviation. The region $\delta\leq \max\delta$ is defined such that $O(\delta^4)$ terms are at most equal to the $O(\delta^2)$ terms.  Therefore within this region, a quartic $\lambda_f(\delta_\tau, g_H^2)$ should reasonably approximate $\lambda(\delta_\tau, g_H^2)$. By performing calculations at three or more $\delta_\tau$ each with $\epsilon_A$, one would expect the value of $|\lambda_f(0, g_H^2) - \lambda(0, g_H^2)|$ and therefore $\epsilon_B$ to be larger than $\epsilon_A$ only by an order-one factor. 
In addition, we will require the condition in  \eq{LooseBound}: $| \lambda_f(\delta,g_H^2) - \lambda(\delta,g_H^2)|\leq M+\epsilon_B$ in the whole complex plane satisfying $|\delta| < \max\delta$. As shown in Appendix~\ref{ap:ACerror proof},  we derive an upper bound on the error of the temporal scale-setting for \textbf{Scheme B}:

\begin{align}\label{eq:SchemeB}
 |\lambda(i\delta_t, g_H^2) -\lambda_f(i\delta_t, g_H^2) |&\lesssim \epsilon_B\left(\frac{M + \epsilon_B}{\epsilon_B}\right)^{ \omega(\delta_t)}
\end{align}
with $0<\omega(\delta_t) = \frac{4}{\pi}\arctan\frac{\delta_t}{\max \delta} < 1$ following the Lemma 1 in \cite{Miller:1970SIAM}.

\begin{figure}
\centering
\includegraphics[width=1.0\linewidth]{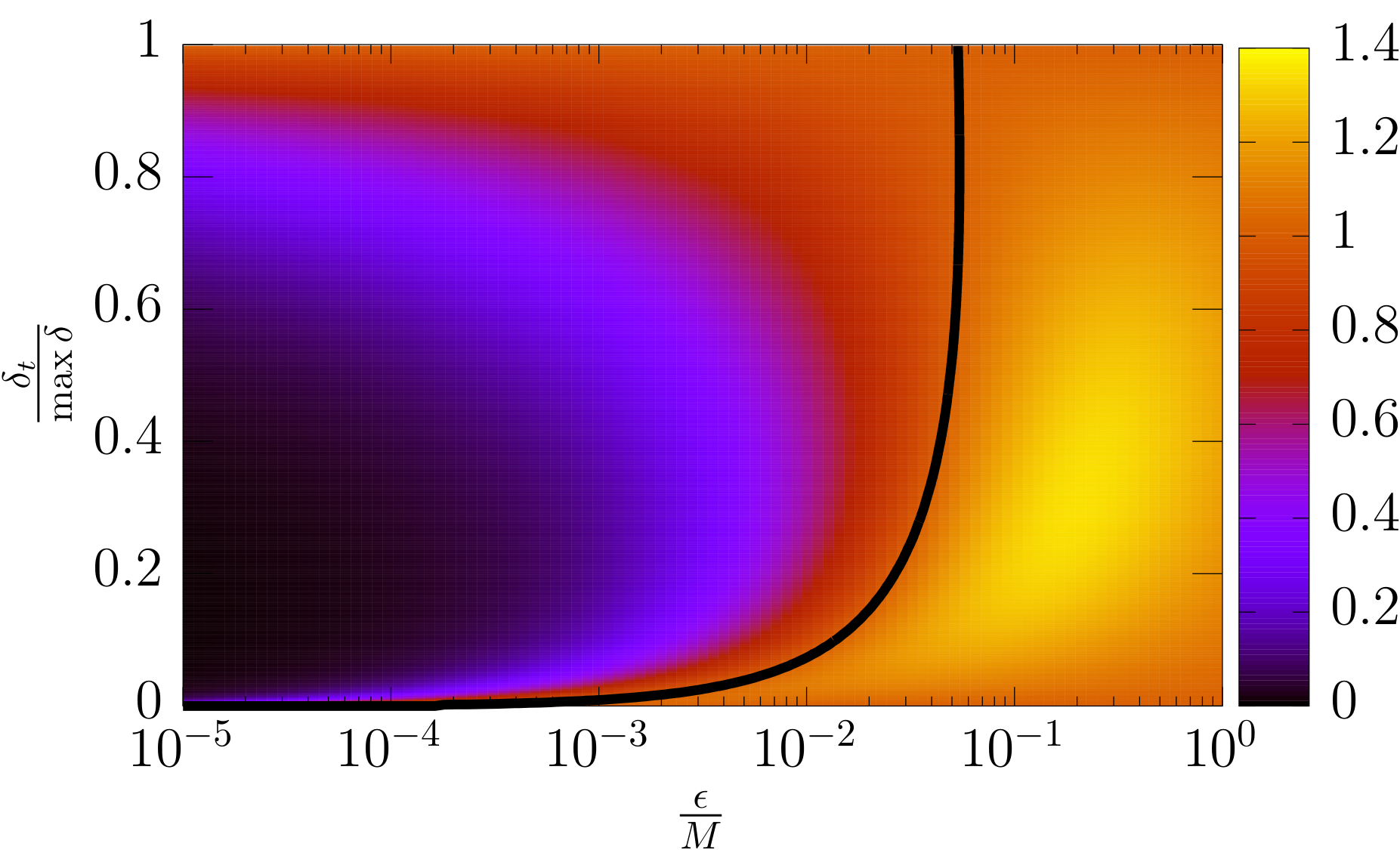}
\caption{\label{fig:error}Ratio of the error bounds of \textbf{Scheme B} in \eq{SchemeB} to that of \textbf{Scheme A} in \eq{SchemeA}. The black line indicates when the bounds are equal.}
\end{figure}

The advantage of using \textbf{Scheme B} could be seen from comparing the ratio of \eq{SchemeB} to \eq{SchemeA} when setting $\epsilon_A = \epsilon_B = \epsilon$.
The error in \textbf{Scheme A} has a quadratic dependence on $\delta_t$. When $\epsilon/M\ll1$, the growth of the errors in \textbf{Scheme B} is delayed with respect to that in \textbf{Scheme A}, until $\delta_t$ is very close to $\max \delta$, as shown in \fig{error}. It results that
such advantage of \textbf{Scheme B} over \textbf{Scheme A} requires  $\epsilon/M < 0.053$, where around $\epsilon/M = 0.053$, the derivatives of the two prefactors respective to $\delta_t$ are the same at $\delta_t = \max \delta$. Ultimately, the overall accuracy of the scale setting in both schemes is controlled by the magnitude of the error bound parameter M.
All the above holds, unless the error bound parameter $M$ is too large such that simulations at very small $\delta_t$ values are necessary to have the overall accuracy under control, which are computationally very expensive.

The result that \textbf{Scheme B} performs better at small $\epsilon/M$ ratio has clear physical interpretations. Correcting BCH errors as 
\textbf{Scheme B} does via analytic continuation becomes more important when the error bound parameter $M$ is larger. On the other hand, in the \textbf{Scheme B}, the analytic continuation is sensitive to the Euclidean precision $\epsilon$ as seen from \eq{SchemeB}, while \textbf{Scheme A} is less affected. For \textbf{Scheme B} to yield a smaller error, certain accuracy of the functional form $\lambda_f(\delta_\tau, g_H^2)$ has to be achieved. This leads to the practical concern that a large amount of resources on the Euclidean calculation might be required at small couplings as the signals become weaker. In addition, for the low energy states involved in the scale-setting procedure, the error bound parameter $M$ could be smaller than the estimation using \eq{delta_max}. Therefore, we expect both \eq{SchemeA} and \eq{SchemeB} to be conservative bounds, as one can confirm by comparing \tab{AB comparison} and \fig{scale}. In such cases, a smaller error $\epsilon$ would be required to obtain the same values of $\epsilon/M$ that govern the suppression of the errors in \textbf{Scheme B}.

In Table \ref{tab:AB comparison} we estimate for $D_4$ models with 2 and 3 spatial dimensions and 
different number of plaquettes/number of links/different values of the bare coupling $g_H$, the largest possible trotter step compatible with 0.1 errors on the temporal-scale-setting for \textbf{Scheme A} and \textbf{Scheme B}, assuming $\epsilon = 0.02$.
We observe that as expected, the advantage of \textbf{Scheme B} to allow the use of relative large trotter step at a given systematic error level is remarkable for larger systems and stronger couplings, that in turn corresponds to larger values of $M$.
 
\begin{table}[ht]
    \centering
    \begin{tabular}{cccc|cccc}
\hline \hline
$d$ & $\mathcal{N}_{\text{plaquette}}$& $\mathcal{N}_{\text{link}}$ & $g_H^2$ &$\max\delta$ &M & $\delta_t^A$ & $\delta_t^B$\\
\hline 
\multirow{3}{*}{2} &\multirow{3}{*}{2}& \multirow{3}{*}{4}  & 0.33 &0.041 & 0.23 & 0.024 & 0.022\\
&&& 0.71  &0.089 & 2.1 & 0.017 & 0.025\\
&&& 1.0 &0.059 & 2.0 &  0.012  & 0.017\\
\hline
\multirow{3}{*}{2} & \multirow{3}{*}{$ 4^2$} & \multirow{3}{*}{$32$}  & 0.1 &0.013 &$5.8\times 10^{-4}$ & 0.013 & 0.013 \\
&&& 0.5 &0.063 &7.2 & 0.0066 & 0.014\\
&&& 1.0  &0.059 &16  & 0.0042  & 0.011\\
\hline
\multirow{3}{*}{3} & \multirow{3}{*}{$ 4^3$}  & \multirow{3}{*}{$192$}  & 0.1 &0.0063 &$3.5\times10^{-3}$ & 0.0063 & 0.0063\\
&&& 0.5  &0.032 &43 & 0.0013 & 0.0052\\
&&& 1.0  &0.059 &192 & 0.0012  & 0.0082\\
\hline\hline
\end{tabular}
    \caption{Benchmark values for $D_4$ models in $2+1D$ and $3+1D$ with periodic conditions. Assuming $\epsilon=0.02$  and demanding the errors in \eq{SchemeA} and \eq{SchemeB} to be below $0.1$, $\delta_t^{A},\delta_t^{B}$ are largest possible $\delta_t$ for $\textbf{Scheme A}$ and $\textbf{Scheme B}$. When $g_H^2=0.1$, $M<0.1$ and we take $\delta_t^{A}=\delta_t^{B}=\max\delta$. The first three rows use the same parameters as in \fig{scale}. The absolute error in \tab{AB comparison} can be converted to the relative error in \fig{scale} by use of a factor $\delta_t / a_t m_{\plaq}\approx 1$ from \tab{fits}.}
    \label{tab:AB comparison}
\end{table}

\section{Numerical Results}
\label{sec:num}
In this section we present a concrete demonstration of some of the theoretical perspectives discussed above by using a two-plaquette theory with the discrete, non-Abelian gauge group $D_4$. We perform classical simulations of a quantum computer using \texttt{qiskit}~\cite{santos2017ibm,cross2017open}. These calculations are performed without modeling of realistic noise sources, corresponding to a perfect, error-free quantum computer.

\begin{figure}
\centering
\includegraphics[width=0.8\linewidth]{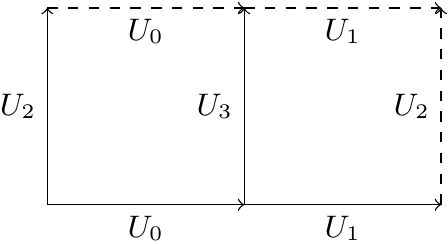}
\caption{\label{fig:lattice}The lattice geometry used for the $D_4$ gauge simulation. The plaquettes are given by $U_2^\dagger U_0^\dagger U_3 U_0$ and $U_3^\dagger U_1^\dagger U_2 U_1$.  Dash lines are used to indicate repeated links due to the periodic boundary conditions.}
\end{figure}

We simulate the $D_4$ gauge field on the two-dimensional lattice shown in \fig{lattice}. This lattice represents the smallest two-dimensional lattice which cannot be reduced to a one-dimensional theory. The simulations requires a five $D_4$ registers, and uses a total of 17 qubits: 12 for physical degrees of freedom, 3 for an ancillary group register, and 2 ancillary qubits. Note that, for brevity, we have broken with the notation of previous sections, in referring to a link not by the source and sink sites, but instead with a single direct index  $i=0\ldots 3$.

We define a trace on $D_4$ (not uniquely specified by the group structure) by embedding $D_4$ into $U(2)$, and defining the trace via the fundamental representation of that Lie group. The embedding of $D_4 < U(2)$ is generated by the elements $\sigma_x$ and $i \sigma_z$. 
The Hamiltonian terms are
\begin{align}
\label{eq:hamid4}
H_V =& -\frac{1}{\delta_t}\log T_V\nonumber \\ =&
-\frac{1}{g^2_H}\bigg(\Re \Tr \left[U_2^\dagger(t) U_0^\dagger(t) U_3(t) U_0(t)\right]
\nonumber\\&+ \Re \Tr \left[U_3^\dagger(t) U_1^\dagger(t) U_2(t) U_1(t)\right]\bigg)
\nonumber\\
H_K = &-\frac{1}{\delta_t}
\sum_{i=0..3} \log T_K^{(i)}
\end{align}
where $\log T_K^{(i)}$ is the one-link kinetic term for the $i$-th link, determined as discussed in Appendix~\ref{eq:distransfer}. \footnote{For the numerical calculations, we take $\delta_t = 1$ in $H_K$ specifically for the discrete group, such that the eigenvalues of the kinetic Hamiltonian are sufficiently differentiated when the system is evolved with small trotterization step. This trajectory is adequate since, using the character expansion for the kinetic part of the transfer matrix in Appendix.~\ref{sec:discrete}, one can show that this construction captures the kinetic energy in the continuous time limit taken for a discrete group.} In total, the quantum simulations entailed $\sim 200$ entangling gates per $\delta$~\cite{Lamm:2019bik}. This is roughly the resources recently used in ~\cite{nam2019ground,Arute:2019zxq,1084}, suggesting that a single step of time evolution may be possible on current quantum devices. 

Stochastic state preparation has been demonstrated for thermal states in $D_4$~\cite{Harmalkar:2020mpd} and particle-like states in $\mathbb{Z}_2$~\cite{Gustafson:2020yfe}. While these results are promising, the initial states are found to have contamination from excited states that complicates the analysis.  Therefore, to simplify the study of trotterization and the continuum limit, we use exact diagonalization of the Kogut-Susskind Hamiltonian to compute the eigenvalues and then construct our initial state as
\begin{equation}
\label{eq:inistate}
    |\psi(0)\rangle=\frac{1}{\sqrt{2}}|\psi_0\rangle+\frac{1}{\sqrt{2}}|\psi_i\rangle
\end{equation}
where $|\psi_i\rangle$ correspond to the $i-$th excited state. By preparing such initial states, the corresponding time-dependent state should be
\begin{equation}
\label{eq:exactinit}
    |\psi(t)\rangle_{KS}=\frac{e^{iE_0 t}}{\sqrt{2}}|\psi_0\rangle+\frac{e^{i E_i t}}{\sqrt{2}}|\psi_i\rangle.
\end{equation}

As pointed out in \cite{Gustafson:2020yfe} at finite $a_t$, trotterization mixes Kogut-Susskind eigenstates through nonzero transition matrix elements $\langle\psi_k|H(a,a_t)|\psi(0)\rangle$:
\begin{eqnarray}
    |\psi(t)\rangle &=& e^{iH(a,a_t) t} |\psi(0)\rangle  \notag\\
    &=& \sum_k e^{iE_k(a_t) t} |\psi_k(a_t)\rangle \langle\psi_k(a_t)|\psi(0)\rangle \notag\\
&=& \sum_k \lambda_k e^{iE_k(a_t) t} |\psi_k(a_t)\rangle
    \label{eq:psit}
\end{eqnarray}
 As $a_t\rightarrow 0$, one can show that 
the time-dependent state in \eq{psit} reduces to \eq{exactinit}. Performing measurements of the operator $\mathcal{O}$ on this state with a quantum computer yields
\begin{align}
\label{eq:fittin}
    &\langle\mathcal{O}(t)\rangle\equiv \langle\psi(t)|\mathcal{O}|\psi(t)\rangle \notag\\= &c_{0}+\sum_{k\neq j}\left[c_{k,j}\cos(a_t(E_j-E_k)l)+s_{k,j}\sin(a_t(E_j-E_k)l)\right]
\end{align}
where $l=[0,N]$ is the integer trotter step. $c_{k, j}, s_{k,j}$ incorporate both the mixing effects entailed in $\lambda_k$ and the matrix elements of $\mathcal{O}$. The $E_k$ here and in the following are eigenvalues of $H(a,a_t)$ and have explicit dependence on $a_t$. This expression, analogous to euclidean lattice field theory operators, can be used to fit the energies of states $a_t m_i\equiv a_t(E_i-E_0)$ where $E_0$ is the ground state energy.

It is useful here to compare the fitting procedure to that of Euclidean LFT.  In Euclidean space, $\langle\mathcal{O}(\tau)\rangle\sim\sum_i \alpha_i e^{-E_i\tau}$ and thus taking $\tau\rightarrow \infty$ acts as a low-pass filter which removes higher energy states.  In this way, for sufficiently large $\tau$, $\langle\mathcal{O}(\tau)\rangle$ should be exponentially dominated by a single state $E_1$ and one can extract $a_0E_1$.  In contrast, $\langle\mathcal{O}(t)\rangle\sim\sum_i \beta_i e^{-iE_it}$ and thus the excited-state contribution to the lattice matrix element doesn't decrease as $t\rightarrow \infty$. This lack of a natural low-pass filter is why real-time evolution can access matrix elements that could be inaccessibly to Euclidean LFT; it also means that the excited-state contamination from imprecise state preparation and trotterization cannot be trivially removed. In this sense, the sampling advantage of quantum computers could be jeopardized by errors induced from excited state contamination. 

In this work, we will use two different spatial Wilson loops as our operators $\mathcal{O}$, which have different quantum numbers, and therefore are sensitive to different eigenstates. The first is the left plaquette, $\mathcal{O}_{\plaq}=\Re\Tr U_0U_3U_0^\dag U_2^\dag$, and by following the construction of \cite{Lamm:2019bik} it can be measured without any ancillary qubit. The second is the Wilson loop over the entire lattice, $\mathcal{O}_{\tuplaq}\sim\Re\Tr U_0U_1U_2U_1^\dag U_0^\dag U_2^\dag$, which requires an ancillary group register to be computed~\cite{Scotboat}. For each of these operators, we construct different initial states from Eq.~(\ref{eq:inistate}). 

\begin{figure*}[t]
\centering
\includegraphics[width=\linewidth]{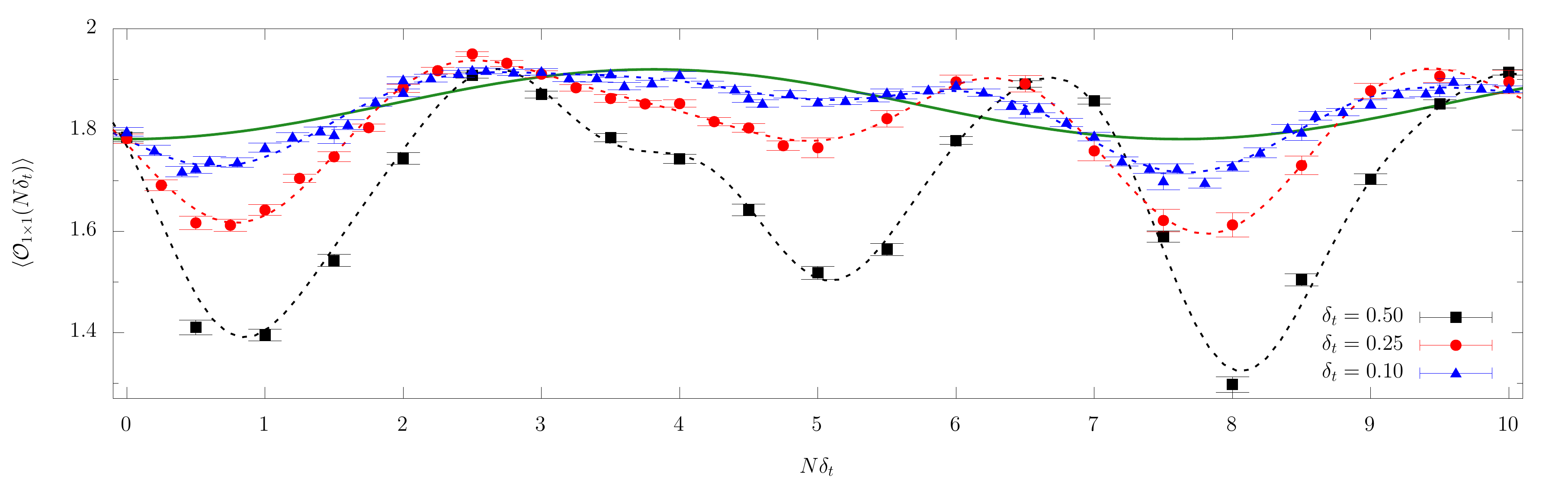}
\caption{\label{fig:trotter_fit} Expectation value of the plaquette $\langle\mathcal{O}_{\plaq}(N\delta_t)\rangle$ vs $N\delta_t$ for different $\delta_t$ with fixed $g_H^2=1.11$. The green line indicated the $\delta_t\rightarrow 0$ exact results}
\end{figure*}

The computations are done for multiple values of $g_H^2(a,a_t)=[0.71,1.25]$ and $\delta_t=[0.01,0.7]$ for $N\delta_t=[10,20]$. 
The circuits used are detailed in~\cite{Lamm:2019bik}. The BCH contributions vanish for $g^2_H(a,a_t)\rightarrow 0$, and as a result, the matrix elements $c_{k,j}, s_{k,j}$ in \eq{fittin} vanish with only $c_{i,0}$ and $s_{i,0}$ surviving. Thus the statistical errors required to resolve oscillations in $\langle\mathcal{O}(t)\rangle$ must be decreased accordingly, calling for increased number of shots. An additional complication from the continuum limit approach is that the gaps $a_t m_j-a_t m_i\rightarrow0$ as $g^2_H\rightarrow 0$ and thus contamination due to trotterization errors grow unless $\delta_t$ is decreased as well. Together, these amount to the cost of the approach to the continuum to scale poorly. For our model, we find that for $g_H^2(a,a_t)>1$, the required number of shots for the \texttt{qiskit} noiseless simulator \texttt{qasm} ranged from 1600 to 64000 as we decreased $\delta_t,g_H^2$.  For this reason,  we utilized \texttt{state\_vector} simulator -- which reports exact probability distributions -- for $g_H^2(a,a_t)\leq1$ in order to investigate the continuum limit at reduced computational cost.  This just emphasizes the importance of being able to perform calculations at large $a,a_t$ on reducing the quantum resources required.
For the $D_4$ theory, the eigenvalues of $H_V$ (\eq{hamid4}) are $1/g^2_H\times\{-4, -2, 0, 2, 4\}$. States evolved under one-step time evolution operator $e^{i \delta_t H_V/2}$ built from $H_V$ will obtain phases in the range $\delta_t/2g^2_H\times[-4,4]$.
For the $\delta_t$ and $g^2_H$ chosen, the phase differences are smaller than $2\pi$ so that one can resolve states with different potential energies.

\subsection{Scale Setting in Minkowski Spacetime}
\label{sec:scalemink}

We evaluate $\langle\mathcal{O}_{1\times 1}(t)\rangle$ to investigate the effect of the renormalization of the temporal scale in Minkowski calculation, by performing fits to Eq.~(\ref{eq:fittin}) and  extracting the lowest energy gap $a_t m_{\plaq}$ from $\langle\mathcal{O}_{\plaq}(t)\rangle$. The initial state is constructed from \eq{inistate} with excited state $i=2$.
An example of these calculations is found in Fig.~\ref{fig:trotter_fit} for fixed $g_H^2(a,a_0) = 1.11$ and three different $\delta_t = \{0.5,0.25, 0.1\}$. Comparing the trotterized results to the continuous-time one calculated using classical computations, the state contamination can be clearly observed and decreases for smaller values of $\delta_t$. The results for $a_t m_{\plaq}$, for the whole range of $g^2_H$ and $\delta_t$ are found in Table~\ref{tab:fits}. We observe that the bare mass $a_t m_{1\times 1}/\delta_t$ is changed by only 3\% when $\delta_t$ increases from 0.1 to 0.5 for $g^2_H = 1.11$, while the mixing effect could be changed by around 20\% from comparing the $c_{k,j}, s_{k,j}$ amplitudes in \fig{trotter_fit}.

\begin{table}[ht]
\caption{Numerical results for lattice masses $a_t m_{\plaq}$ and $a_t m_{\tuplaq}$ for the bare couplings $g_H^2$ and $\delta_t$ studied here. Rows above (below) the line indicate  \texttt{qasm} (\texttt{state\_vector}) results.}

\begin{tabular}{cccc}
\hline\hline
$g_H^2$&$\delta_t$&$a_t m_{\plaq}$ &$a_t m_{\tuplaq}$\\
\hline
1.25& 0.70 & 0.6663(11) & ---\\
1.25& 0.65 & 0.606(17) & ---\\
1.25& 0.60 & 0.554(16) & ---\\
1.25&0.50 & 0.442(6) & 1.120(15)\\
1.25& 0.40 & 0.349(7) & ---\\
1.25&0.25 & 0.211(4) &0.575(5)\\
1.25&0.10 & 0.0821(11) &0.2332(11)\\
1.25& 0.05 & 0.0414(5) & ---\\
1.25& 0.01 & 0.00819(11) & ---\\
1.18 & 0.5 & 0.446(13) &1.12(4)\\
1.18 & 0.25 & 0.207(3) &0.557(9)\\
1.18 & 0.1 & 0.0838(19) &0.223(3)\\
1.11 & 0.5 & 0.429(11) &1.016(11)\\
1.11 & 0.25 & 0.206(10) &0.518(11)\\
1.11 & 0.1 & 0.0832(7) &0.208(4)\\
1.05 & 0.5 & 0.404(14) &0.918(11)\\
1.05 & 0.25 & 0.1987(4) &0.478(5)\\
1.05 & 0.1 & 0.08103(14) &0.189(3)\\
\hline
1 & 0.1    & 0.07580(8)     &0.1749 (7)\\
1 & 0.05   & 0.037855(14)   &0.0883 (4)\\
1 & 0.01   & 0.007584(4)    &0.017572(6)\\
0.91 & 0.1 & 0.06699(9)     &0.1474(4)\\
0.91 & 0.05& 0.03348(3)     &0.07402(9)\\
0.91 & 0.01& 0.0066953(14)  &0.014835(4)\\
0.83 & 0.1 & 0.0575(4)      &0.1250(6)\\
0.83 & 0.05& 0.02889(10)    &0.06266(16)\\
0.83 & 0.01& 0.005791(8)    &0.012493(14)\\
0.77 & 0.1 & 0.0488(8)      &0.1051(8)\\
0.77 & 0.05& 0.02462(19)    &0.05230(15)\\
0.77 & 0.01& 0.004948(8)    &0.01049(3)\\
0.71 & 0.1 & 0.0424(3)      &0.0885(3)\\
0.71 & 0.05& 0.02102(9)     &0.04387(7)\\
0.71 & 0.01& 0.004191(9)    &0.008753(7)\\
\hline\hline
\end{tabular}
\label{tab:fits}

\end{table}

\begin{figure}
\centering
\includegraphics[width=\linewidth]{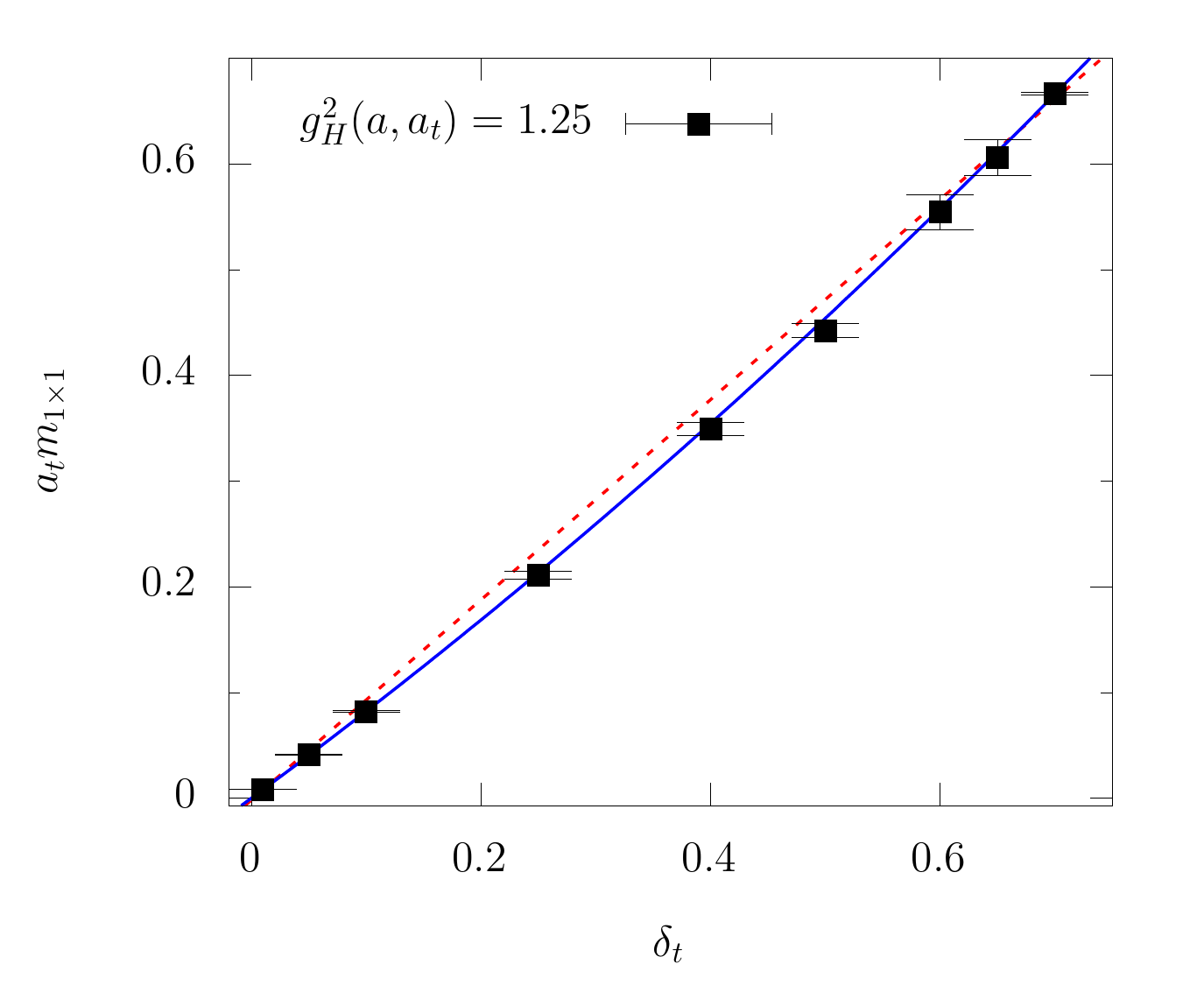}
\caption{\label{fig:at_fit} $a_t m_{\plaq}$ vs $\delta_t$ for $g_H^2=1.25$. The dashed (solid) lines indicate a linear (quadratic) fit to the data. The poor linear fit ($\chi^2/{\rm d.o.f} = 40.2$) demonstrates that $a_t$ is not proportional to $\delta_t$ as might be naively expected. $\chi^2/{\rm d.o.f} = 1.1$ for the quadratic fit.}
\end{figure}

\begin{figure}
\centering
\includegraphics[width=\linewidth]{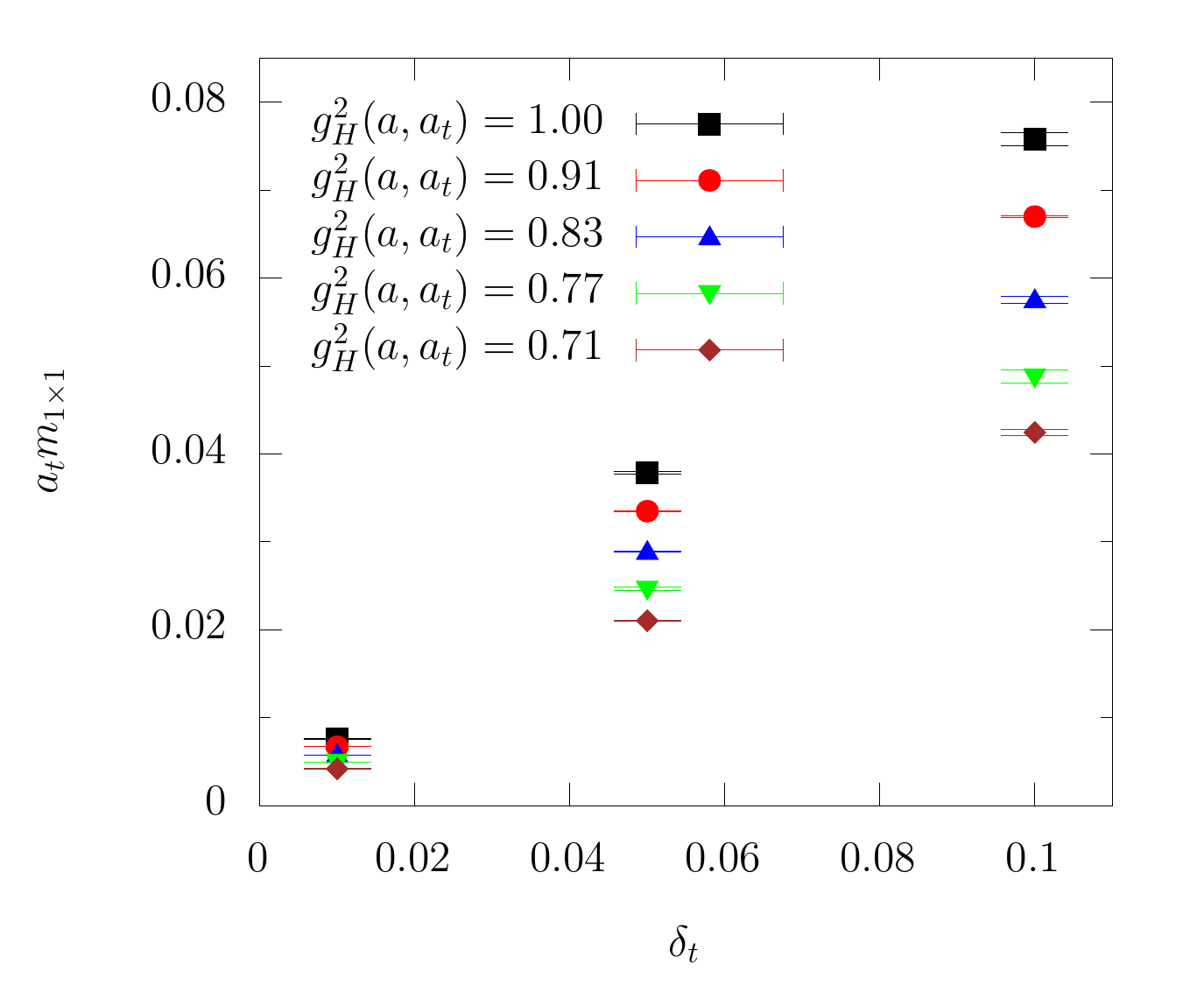}
\caption{\label{fig:at_fit2} $a_t m_{\plaq}$ vs $\delta_t$ for different $g_H^2$.  Notice that renormalization of $a_t$ is clearly $g_H^2$ dependent.}
\end{figure}

In the Euclidean lattice calculation, using the relation in \eq{renorm} obtained in the weak coupling limit~\cite{Hasenfratz:1981tw,Karsch:1982ve,Hamer:1996ub}, the perturbative renormalized anisotropy is
\begin{align}
\label{eq:xivt}
    \xi^{-1}=&~c(a,a_0)\delta_ \tau\sim\left[\frac{1-c_s(\xi)g_H^2}{1-\frac{c_t(\xi)+c_s(\xi)}{2}g_H^2}\right]\delta_\tau\nonumber\\\approx&\left[1+\frac{c_t(\xi)-c_s(\xi)}{2}g_H^2(a,a_0) + O(g^4_H)\right]\delta_\tau.
\end{align}
Even without a functional form for $c_i(\xi)$, one can see that solving Eq.~(\ref{eq:xivt}) self-consistently for $\xi$ will break $a_0\propto\delta_\tau$.
For $SU(N)$ in 3 and 4 dimensions, $c_{+}(\xi)\equiv c_t(\xi)+c_s(\xi)\approx O(10^{-2})$ and $c_{-}(\xi)\equiv c_t(\xi)-c_s(\xi)\approx O(10^{-1})$ ~\cite{Hasenfratz:1981tw,Karsch:1982ve,Hamer:1996ub} -- suggesting that the nonlinear renormalization is generically small. While the exact functions $g_H^2(a,a_0),c_i(\xi)$, depend upon the metric, the form of \eq{xivt} should remain unchanged. It is possible to investigate Eq.~(\ref{eq:xivt}) in Minkowski space by rewriting it as $a_t=c(a,a_t)a\delta_t$. For the case of $g_H^2=1.25$ plotted in Fig.~\ref{fig:at_fit}, one sees that $a_t$ has a clear $\delta_t^2$ dependence. Furthermore, Fig.~\ref{fig:at_fit2} shows that the linear dependence of $a_t$ upon $\delta_t$ is a $g_H^2$ dependent function. Together, these demonstrate the renormalization of $a_t$ with respect to the bare coupling $g_H$ and the trotter step $\delta_t$. Although for our toy model with two spatial plaquettes we cannot extract the spatial scale $a$ from numerical results, we can relate $a$ to the running of $g$ using the perturbative results~\cite{Christou:1998ws}
\begin{equation}
\label{eq:running}
    a \Lambda=\exp\bigg(-\frac{1}{2b_0 g^2}\bigg)(b_0 g^2)^{-\frac{b_1}{2b_0}}(1+O(g^2)).
\end{equation}
In the above, $b_0,b_1$ are the standard $g^3,g^5$ coefficients in the perturbative $\beta$ function.
Altogether, to extract physical properties from quantum simulations, we must first determine both scales explicitly. 

\subsection{Comparison of Schemes for Scale Setting}
\label{sec:scaleeucl}
To reduce the quantum resources for lattice calculation, we determine the scale using the Euclidean data following the procedure proposed in Sec.~\ref{sec:mlft}. As a demonstration, we will focus on the temporal lattice spacing and show its determination using \textbf{Scheme A} and \textbf{Scheme B}. 

\begin{figure*}[t]
\centering
\includegraphics[width=0.48\linewidth]{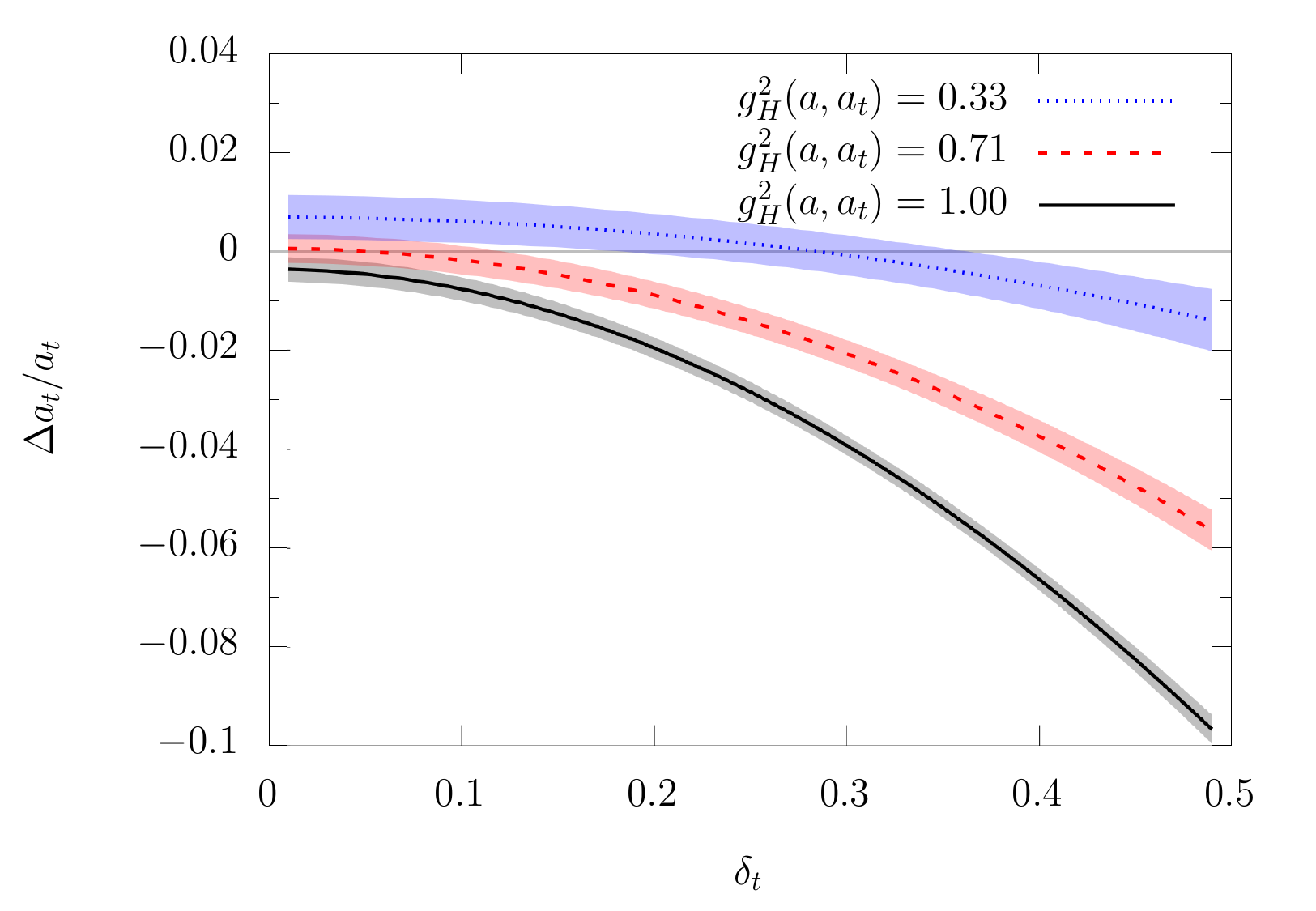}
\includegraphics[width=0.48\linewidth]{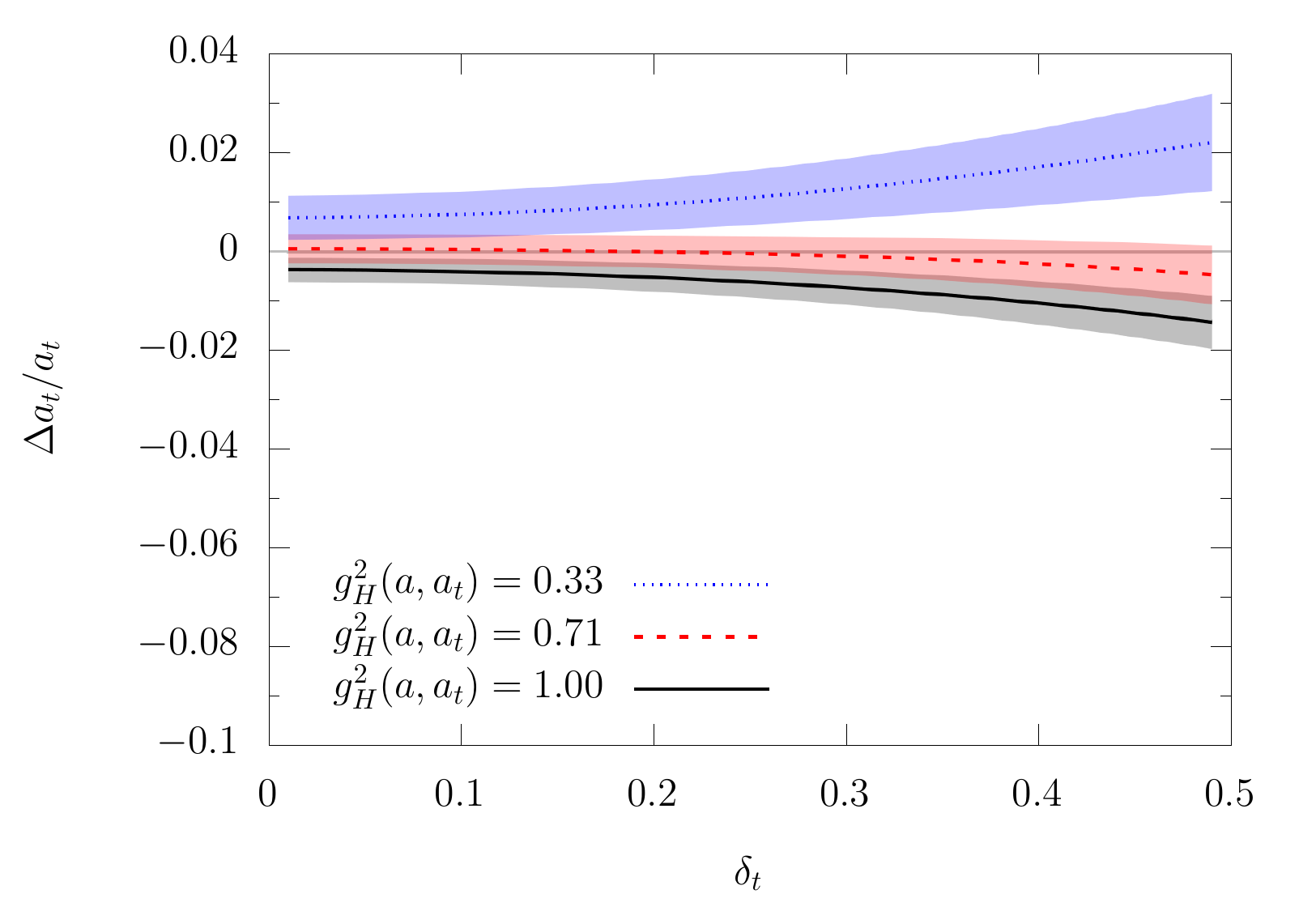}
\caption{\label{fig:scale} Errors with respect to the truth scale in a Minkowski calculation using \textbf{Scheme A} (left) and \textbf{Scheme B} (right), with the band showing the error from the $1\sigma$ region in the Euclidean fitting. We have assumed that the measurement precision in Euclidean spacetime could reach $1\%$. Together with fitting errors, we have $\epsilon \le 2\%$ at 68\% C.L. for the $g^2_H$ and $\delta_\tau = \delta_t$ used in this figure. For \textbf{Scheme A} we use $\Delta a_t$ from the fit, as $\epsilon$ is barely affected by the fitting procedure.}
\end{figure*}

We solve the eigenvalues of the Euclidean transfer matrix built from $H_K$ and $H_V$ in \eq{hamid4}. By taking the logarithm of these eigenvalues, we could extract the energies of the spectra on a Euclidean space-time, with the ground state energy normalized to zero. The observable $a_0 m_{1\times 1}$ is the energy corresponding to the 2nd excited state. With the set of bare couplings $1/g^2_H = [0.8, 3.0]$ and $\delta_\tau = [0.5, 0.01]$, we obtain $a_0 m_{1\times 1}$ and in addition we apply an error of $1\%$ to $a_0 m_{1\times 1}$ for any bare coupling and $a_0$ used, in order to mimic the measurement error using classical Monte Carlo simulations.
Then we do a 2D fit to get the functional dependence of $a_0 m_{1\times 1}$ on both $\delta_\tau, g^2_H$ as:
\begin{eqnarray}
\label{eq:scalefit}
a_0 m_{1\times 1}(\delta_\tau, g^2_H) &=& \delta_\tau \bigg(f_1(g^2_H) + f_2(g^2_H) \delta^2_\tau\bigg)
\end{eqnarray}
Given that spatial lattice spacing $a$ is related to $g^2_H$ via the relation in \eq{running} and $f_i$ functions should be power-law expansions of $a$, we use the following functional form for the 2D fitting of $a_0 m_{1\times 1}$
\begin{equation}
\label{eq:ffun}
    f_i(g^2_H) =\sum_{n} a_{i,n}\bigg(\exp(-\kappa_0/g^2_H)\frac{1}{g^{2 \kappa_1}_H}\bigg)^n.
\end{equation}
The fitted $a_0 m_{1\times 1}$ ($\chi^2/{\rm d.o.f} = 1.3$) is at most 3\% away from their truth value at 68\% C.L.. 
With this, we can then analytically continue with either scheme. In \fig{scale}, the percent systematic error for both schemes compared to $a_t$ computed by diagonalizing $U(a_t)$ in Sec.~\ref{sec:scalemink} (which agree with \tab{fits}). The colored region shows the errors from the $1\sigma$ band in the Euclidean fitting. For the three $g_H^2$ used in \fig{scale}, the errors $\epsilon\leq2\%$ at 68\% C.L., with a weak dependence on $\delta_\tau$. 

As expected, $\mathbf{Scheme~A, B}$ give precise evaluations of $a_t$ at small $\delta_t$. When $\delta_t$ gets larger, the BCH errors of \textbf{Scheme A} should increase as $\delta_t^2$.  BCH errors get smaller for smaller $g^2_H$, as which lowers the error bound parameter $M$. As shown in \fig{scale}, for $g^2_H \leq 0.33$($\epsilon\sim M$), \textbf{Scheme A}  provides a  better estimation of the lattice scales for the whole region of  $\delta_t < 0.5$.
The errors from the analytical continuation of \textbf{Scheme B} barely increase for larger values of $\delta_t$, and are in principle less sensitive to $g^2_H$ as expected. For the assumed precision in the Euclidean data and the fitting procedure,
the error, $\epsilon$, is slightly larger for the small $g^2_H$ region and altogether a delayed growth in the errors is maintained for $g^2_H \geq 0.71$. One can conclude that if $g^2_H $ is not particularly small, \textbf{Scheme B}  is more likely to render the lowest errors in the scale determination. In both cases, the observed $2\%$ errors are well within \eq{SchemeA} and \eq{SchemeB}, which predict a $10\%$ error bound for $\delta_t\leq \delta_t^{A(B)}$ shown in \tab{AB comparison}.

Once the measurement of the renormalized spatial lattice spacing is allowed when considering sufficiently large systems, 
one can obtain the renormalized spatial lattice spacing similarly using only Euclidean data.

\subsection{Approaching the Continuum}
In Euclidean LFT, one uses $a,a_0$ for extrapolating to the continuum limit.  This is because performing extrapolations in terms of the bare parameters requires either very higher order fit functions (and therefore many calculations) or calculations with very small $g_H$ and $\delta_\tau$ (at large computational cost). Instead, using $a,a_0$ is an extrapolation in nonperturbative variables, so the bare couplings can be much larger and the fit functions simpler. The same should be true in Minkowski spacetime. In principle, one can approach the continuum along any trajectory as a function of $a,a_t$, but experience from Euclidean LFT~\cite{Klassen:1998ua, Borsanyi:2012zr}, suggests that taking a trajectory of fixed-$\xi_t$ would smoothly and efficiently extrapolate to the continuum limit for Minkowski LFT. In approximations to $U(t)$ other than trotterization~\cite{PhysRevLett.123.070503,cirstoiu2020variational,gibbs2021longtime,yao2020adaptive}, definitions for a nonperturbative scale like $a_t$ are currently unknown, making the extrapolation to the continuum nontrivial. 

In the continuum limit $a, a_t\rightarrow 0$ and thus $a_t m_i$ vanishes. One must instead explore finite physical quantities such as $a_t m_i/a_t m_{1\times 1}$. Due to the smallness of our lattice, the only states which mix with $|\psi_2\rangle$ are from cutoff effects which have $a_t m_i/a_t m_{1\times 1}\rightarrow\infty$. Therefore, we perform another fit to $\langle\mathcal{O}_{\tuplaq} (t)\rangle$ from which we extract a second physical mass, $a_t m_{\tuplaq}$. The initial state for this purpose is constructed with $i=13$. With these, we can study different approaches to the continuum limit. As the scales set from real-time evolution are already available (\tab{fits}) and \textbf{Scheme B} gives comparable precision, in the following, we use $a_t$ from \tab{fits}. We point out that the precision of the spacings extracted from real-time evolution would be much worse when noise is included.

\begin{figure}
\centering
\includegraphics[width=\linewidth]{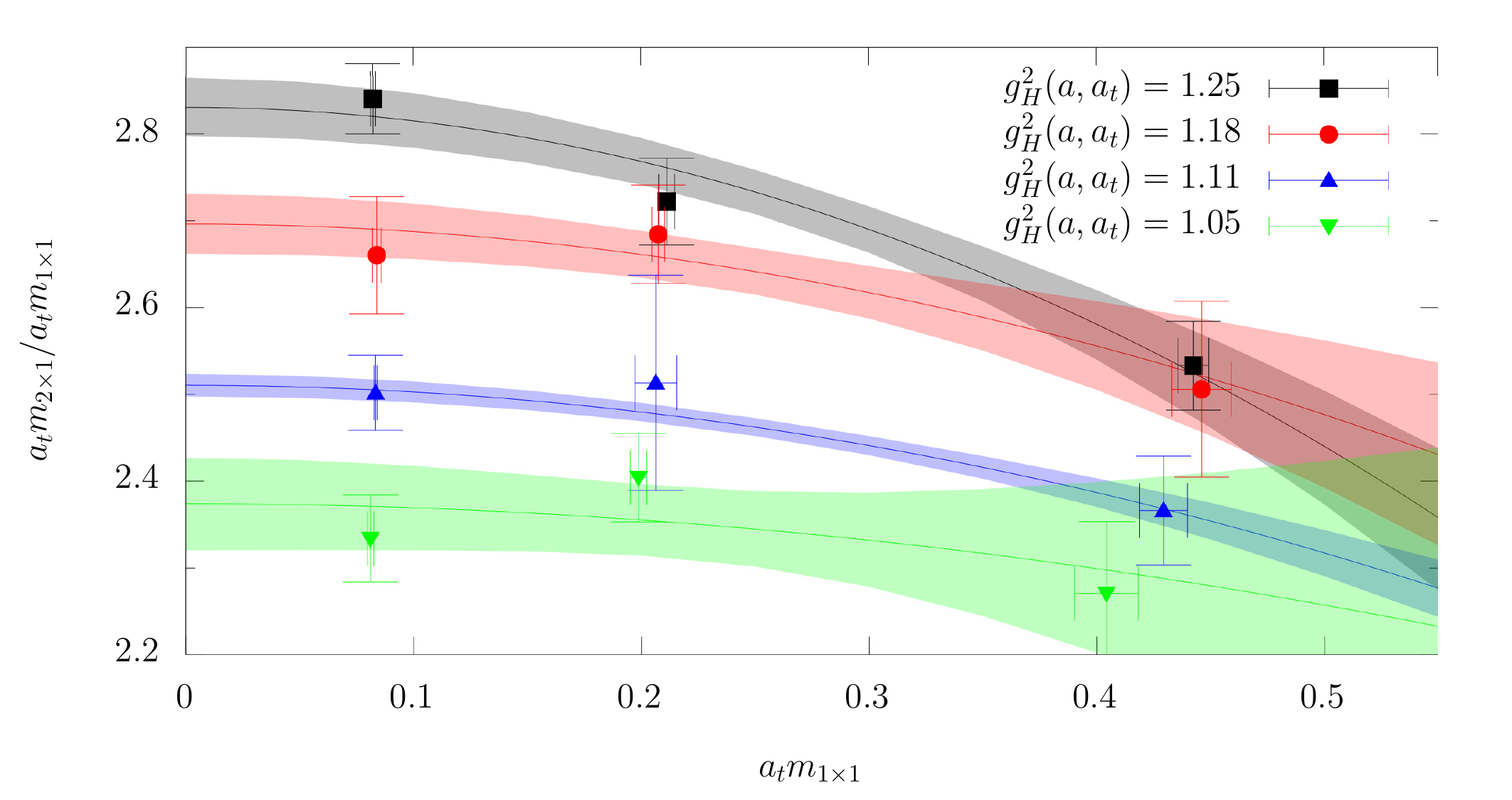}
\caption{\label{fig:mrvat1}$a_tm_{\tuplaq}/a_tm_{\plaq}$ as a function of $a_tm_{\plaq}$ for a variety of $g_H^2(a,a_t)$. The data points are extracted from fixed $\delta_t$ results from a \texttt{qasm} calculation. For each $g^2_H$, we have $\delta_t = \{0.1,0.25,0.5\}$ from left to right. The solid lines reflect the best fit for a fixed $g_H^2(a,a_t)$. The colored bands are the $1\sigma$ uncertainties on the fits.} 
\end{figure}

\begin{figure}
\centering
\includegraphics[width=\linewidth]{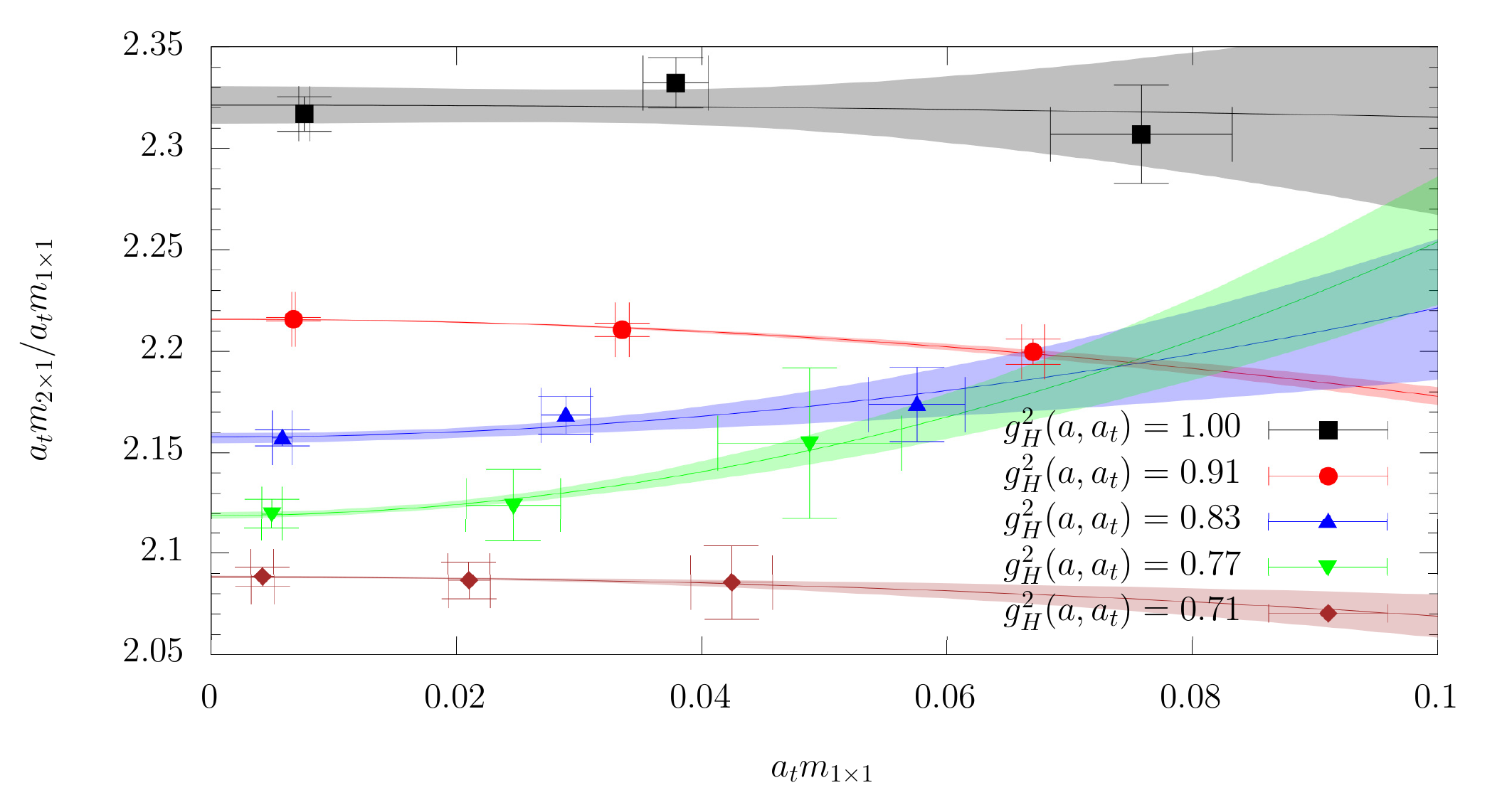}
\caption{\label{fig:mrvat2}$a_tm_{\tuplaq}/a_tm_{\plaq}$ as a function of $a_tm_{\plaq}$ for a variety of $g_H^2(a,a_t)$.  The data points are extracted from fixed $\delta_t$ results from a \texttt{state\_vector} calculation. For each $g^2_H$, we have $\delta_t = \{0.01,0.05,0.1\}$ from left to right. The solid lines reflect the best fit for a fixed $g_H^2(a,a_t)$.  The colored bands are the $1\sigma$ uncertainties on the fits.}
\end{figure}

Numerically, the first trajectory to approach the continuum is to first take $a_t\rightarrow 0$ (the Hamiltonian limit) then take $a\rightarrow0$. Since the lattice spacing errors of $H(a,a_t)$ are $O(a^2,a_t^2)$, we fit the mass ratio to
\begin{equation}
\label{eq:a0fit}
 \frac{a_t m_{\tuplaq}}{a_t m_{\plaq}}=\kappa_0+\kappa_2(a_t m_{\plaq})^2
\end{equation}
 In Fig.~\ref{fig:mrvat1} and Fig.~\ref{fig:mrvat2}, we show these best fit lines of \eq{a0fit} to the data points extracted from \texttt{qasm} and \texttt{state\_vector}. The values of $\kappa_0$ correspond to the Hamiltonian limit value of the mass ratio, and we tabulate them in Table~\ref{tab:fits_ham}. We find good agreement between the $\kappa_0$ in \tab{fits_ham} and those calculated by direct diagonalization of the Hamiltonian with $\delta_t = 0$ for different $g^2_H$.

From Fig.~\ref{fig:mrvat1}, we could see that the difference for $a_t m_{\tuplaq}/a_t m_{\plaq}$ is only 8\% between its value at $\delta_t=0.5$ and the Hamiltonian limit for $g^2_H = 1.11$, while from \fig{trotter_fit}, for the same value of $g^2_H$ and $\delta_t$, the $\langle\mathcal{O}_{1\times 1}(t)\rangle$ could differ from the Hamiltonian limit by up to 20\%. The deviation in $a_t m_{\tuplaq}/a_t m_{\plaq}$ represents BCH corrections to the eigenvalues and the latter one is the correction to wavefunction. Thus less resources (larger trotterization step) could be needed to control the error of $a_t m_{\tuplaq}/a_t m_{\plaq}$ below certain threshold.

Assuming errors scaling as $O(a^2)$ and using \eq{running}, we can fit the $\kappa_0$ in \tab{fits_ham} to the continuum limit using
\begin{equation}
\label{eq:expfit}
    \frac{a_t m_{\tuplaq}}{a_t m_{\plaq}}\bigg|_{a_t\rightarrow0}=\lambda_0+\lambda_1 e^{-\lambda_2 /g_H^2}(g_H)^{\lambda_3}.
\end{equation}

\begin{figure*}[t]
\centering
\includegraphics[width=0.8\linewidth]{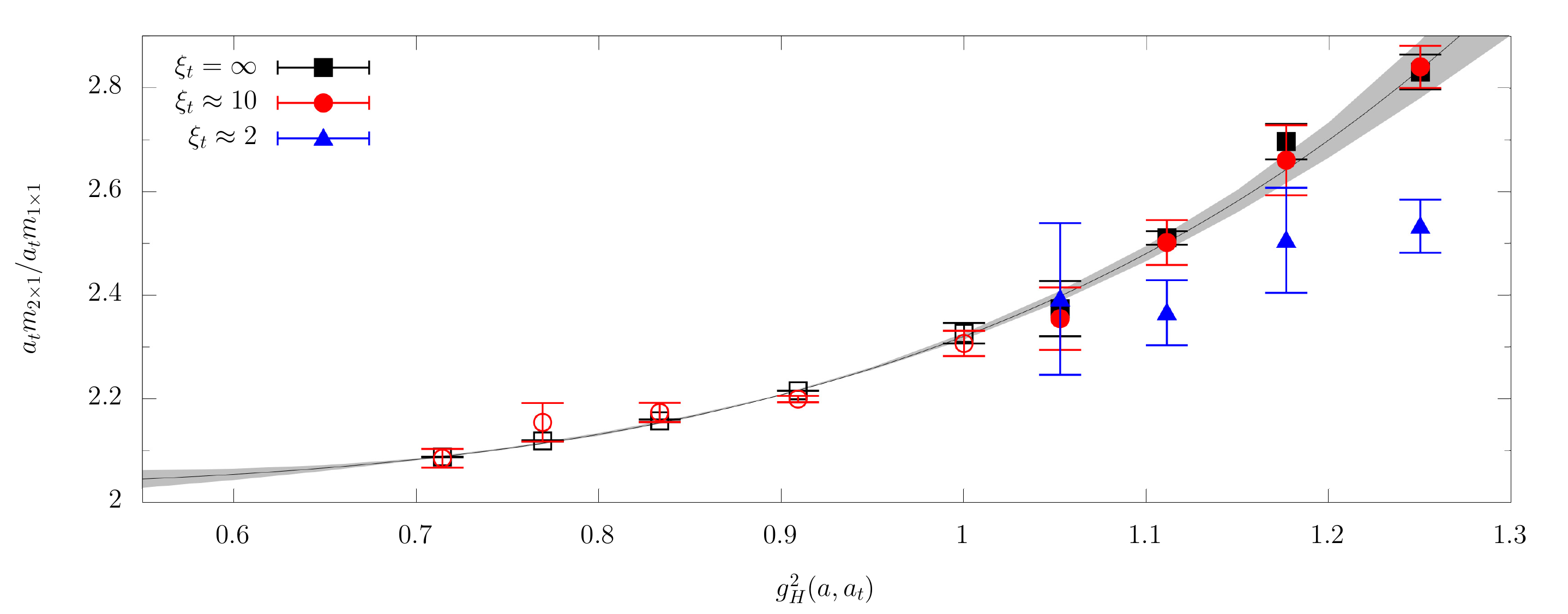}
\caption{\label{fig:mrvg}$a m_2/a m_1$ vs $g_H^2(a,a_t)$ for three continuum trajectories where $\xi_t$ is approximately fixed. Closed and open symbols indicate \texttt{qasm} and \texttt{state\_vector} results respectively. The black line is a best fit to (\protect\blacksq, \protect\blackosq) with a $1\sigma$ uncertainty band.}
\end{figure*}

In contrast to extrapolating in $a_t$, the reliability of this extrapolation depends much more sensitively on being in the perturbative regime, otherwise Eq.~(\ref{eq:running}) receives large corrections and our functional form will be inadequate.  So from \eq{expfit}, we must perform the nonlinear fit in $g_H^2$ which introduces complicated correlations between the fitting parameters. Despite these limitations, we find that the continuum limit result of $\lambda_0=2.03(4)$ agrees with the continuum value $m_{\tuplaq}/m_{\plaq}=2$ computed from the direct diagonalization of the Hamiltonian with $g^2_H\sim 0$ and $\delta_t = 0$. The mass ratio in the continuous time limit, $\kappa_0$ in \tab{fits_ham} (black points), and the best fit line using Eq.~\ref{eq:expfit} are found in Fig.~\ref{fig:mrvg} (black line and grey shaded region for 68\% C.L..).
\begin{table}[ht]
\caption{Numerical results for the dimensionless ratio of lattice masses $a_t m_{\tuplaq}/a_t m_{\plaq}$ after extrapolating for fixed $g_H^2$ to the Hamiltonian limit. Rows above (below) the line indicate  \texttt{qasm} (\texttt{state\_vector}) results.}

\begin{tabular}{cc}
\hline\hline
$g_H^2$&$\frac{a_t m_{\tuplaq}}{a_t m_{\plaq}}|_{a_t\rightarrow 0}$\\
\hline
1.25& 2.83(4)\\
1.185& 2.70(4)\\
1.11& 2.511(13)\\
1.05& 2.37(6)\\
\hline
1.00& 2.321(9)\\
0.91& 2.2159(3)\\
0.83& 2.158(3)\\
0.77& 2.1191(15)\\
0.71& 2.0882(5)\\
\hline\hline
\end{tabular}
\label{tab:fits_ham}

\end{table}

The number of trotter steps $N$ (and therefore gate costs) is proportional to $\delta_t^{-1}$. Clearly therefore, it is an undesirable feature to first take the Hamiltonian limit $a_t\rightarrow 0$ ($\delta_t\rightarrow 0$) and then the continuum limit $a\rightarrow0$. Instead of working at fixed $a$, while decreasing $\delta_t$, one can work at fixed $\xi_t = \frac{a}{a_t}$ and then one could directly approach the continuum limit $a,a_t\rightarrow 0$ through this trajectory. Up to $a^2$ order, the extrapolation in terms of the bare coupling, would be of analogous functional form as that on the right-hand of \eq{expfit}, with the finite-$\xi_t$ effects entering the fitting parameters.
It is thus important to explore how to define the finite $\xi_t$ trajectory. One could perform this tuning on the quantum computer. This would proceed by computing both $a_t,a$ or $\xi_t$ using some physical observables (e.g. Wilson loops or dispersion relations) and then adjust the bare parameters until one has a fixed $\xi_t$ trajectory. This procedure requires multiple quantum simulations to be performed -- albeit at larger $\delta_t$ then an extrapolation to the Hamiltonian limit requires and only the scale-setting observable needs to be computed. But, as demonstrated, we can reliably extract the dependence of the scales $a, a_t$ on $\delta_t$ and $g^2_H$ by analytical continuation using \textbf{Scheme B}. We can then directly invert these relations to find the $\delta_t$ and $g^2_H$ for a fixed $\xi_t$ avoiding the quantum computer entirely for scale-setting.

Given the smallness of our system, we are unable to extract $a$ or the corresponding $\xi_t$ to determine a set of $\delta_t$ and $g^2_H$. This forces us to consider some approximations. While the previous discussions demonstrate renormalization, as the nonperturbative renormalization effects partially cancel out~\cite{Karsch:1982ve}, we expect that the renormalization of $\xi_t$ is milder than that of $a,a_t$ individually. Therefore, we can approximate the fixed $\xi_t$ trajectory as fixed $\delta_t$. In \fig{mrvg}, we show altogether the extrapolation to the continuum limit through the Hamiltonian limit (gray line and black dots), as well as two fixed values of $\xi_t\approx \{10,2\}$ to highlights the power of fixed $\xi_t$ trajectories in achieving the continuum limit results. Indeed, following the trajectory of $\xi_t \approx 10$ (red circle), we can successfully reach the correct continuum limit with clear advantages over the Hamiltonian extrapolation method. For $\xi_t \approx 10$, the needed number of data point simulations is reduced by 67\% when comparing to the Hamiltonian limit procedure that requires all the measurements shown in \fig{mrvat1} and \fig{mrvat2}. In addition, the circuit depth is greatly reduced through avoiding small trotterization steps, e.g. $\delta_t < 0.1$. In \fig {mrvg}, we further investigate using an even smaller $\xi_t \approx 2$, hence $\delta_t = 0.5$ (blue triangles), which implies a further improvement on the circuits depth demands. 
Due to the large uncertainties from quantum simulations as represented by the large error bars from \fig {mrvg}, we calculate the $\frac{a_t m_{2\times 1}}{a_t m_{1\times 1}}$ through explicit diagonalization of the time evolution operator at fixed $\xi_t \approx 1/\delta_t = 2$. We find that the continuum limit is also successfully reached for this large value of trotter step. Although multiple simulations are needed to control the measurement uncertainties on a quantum computer, this highlights the advantages of fixed-$\xi$ calculations in avoiding larger errors from enhanced circuit depth.

\section{Conclusions}
\label{sec:concl}
The natural construction of quantum field theories on quantum computers is as a lattice-regularized theory. Any meaningful calculation must therefore be renormalized and taken to the continuum limit before comparison to experiments can be made. As we have discussed, this involves many hurdles and it will not be easy. Typically quantum simulations are constructed within the Hamiltonian formalism where a spatial lattice with spacing $a$ is time-evolved. Further approximations are required, as the time evolution operator $\mathcal{U}(t)$ built from the Hamiltonian usually cannot be efficiently implemented. 

In this article, we have tackled three important issues related to the simulations of Minkowksi lattice field theories on quantum computers. Firstly, we discussed how renormalization in the form of a temporal lattice spacing $a_t$ arises from trotterizating $\mathcal{U}(t)$.
Secondly, by relating this trotterized time evolution operator to the Euclidean transfer matrix, we propose two different schemes using analytical continuation to set the Minkowski lattice spacings. The most straightforward scheme is to simply equate the Euclidean lattice spacings to those in the real-time calculation (Scheme A). This scheme introduces  $O(\delta_t^2)$ errors in addition to the statistical errors in the Euclidean calculations. 
Our second method (Scheme B) can correct the $O(\delta^2_t)$ errors of Scheme A by analytically continuing the best-fit function of the Euclidean spacings. We have derived conservative bounds on the systematic errors for these schemes for small $\delta_t$, and a loose constraint on the error of the fit to take advantage of Scheme B.
This enables us to reduce the requirements for quantum resources in the scale setting procedure. Thirdly, we show that by taking a fixed anisotropic-$\xi_t$ approach to the continuum limit, one can further reduce the number of simulations with the added benefit of shallower circuits and lower required gate fidelities. We demonstrated these ideas for a $2+1$D, discrete non-Abelian $D_4$ model using \texttt{qiskit} noiseless simulators. 

The results of this work suggest a number of followups.
In particular, these procedures could be tested in the near-term for $\mathbb{Z}_2$ gauge theories in 2+1 dimensions on multiple lattice sizes, allowing for the incorporation of finite-volume effects and the explicit calculation of $a$. Additionally, since the quantum resources increase as $a_t,a\rightarrow 0$, improved Hamiltonians that account for both lattice effects and quantum noise could accelerate extrapolation to the continuum approach while reducing the quantum resources required. Finally, extending the error bounds on our scale-setting schemes to larger $\delta_t$ would be well motivated to ensure that the
systematic errors introduced are under control.  

\begin{acknowledgements}
We would like to thank Joseph Lykken, Judah Unmuth-Yockey, and Michael Wagman for insightful discussions and comments on the manuscript. This work is supported by the Department of Energy through the Fermilab QuantiSED program in the area of "Intersections of QIS and Theoretical Particle Physics". Fermilab is operated by Fermi Research Alliance, LLC under contract number DE-AC02-07CH11359 with the United States Department of Energy. 
\end{acknowledgements}

\appendix

\section{Continuous limit for discrete group with Wilson action}
\label{sec:discrete}
In this appendix, we will show the subtleties in taking the Hamiltonian limit $a_0\rightarrow 0$ constructed from the Wilson action of a discrete gauge group, relying on the spectra of the kinetic part of the transfer matrix, which could be solved using the character expansion - Fourier transformation to the character basis. Taking $D_n$ group as an example (with $n$ even), its irreducible representations $\rho_r$ are as follow:
\begin{eqnarray}
\rho_{1, 1}(j,m)&:& 1~~~~~~~~~~ \rho_{1, 2}(j,m): (-1)^m\\\notag \rho_{1, 3}(j,m)&:& (-1)^j~~~~ \rho_{1,4}(j,m):(-1)^{m+j}\\ \notag
\rho_{2, k}(j,m)&:&
\exp\bigg(i \frac{2\pi j}{n}k\sigma_z\bigg)\sigma^m_x
\end{eqnarray}
with $j\in [0,...,n-1], m\in\{0,1\}$ defining the elements of the group and $k\in\{ 1,2,...\frac{n}{2}-1$\} indexing the different dimension-2 irreducible representations.

The kinetic part of transfer matrix is constructed in the faithful irreducible representation of $D_n$: $\rho_{2,1}(j, m)$, to ensure its positivity~\cite{Lamm:2019bik}. Consider a single link, say the first link $U_1$, it takes values of $\rho_{2,1}(g_i)\equiv\rho_{2,1}(j, m) $, with $i=1$ representing the identity group element of $j=m=0$. Using character expansion, the kinetic part of the transfer matrix for the first link $T^{1}_K$ can be explicitly written as:
\begin{eqnarray}
\label{eq:distransfer}
    T^{1}_K(i,i') &=& \exp\bigg(\beta_t \Re\Tr\left[\rho_{2,1}^\dagger(g_i) \rho_{2,1}(g_{i'})\right]\bigg) \\\notag
    &=& \sum_r d_r c_r \Tr[\rho_{2,1}^\dagger(g_i) \rho_{2,1}(g_{i'})],
\end{eqnarray}
with $d_r$ the dimension of the $r$ representation. The coefficients for the modes in the irreducible representation $r$ are given by $c_r =\sum^{2n}_{i=1} \frac{1}{d_r}\Tr[\rho_r(g^{\dagger}_{i})] T^{1}_K(i, 1)$, explicitly as:
\begin{align}
\label{eq:coefficient}
    c_{1,1} &= \sum^{n-1}_{j=0} (e^{2\beta_t \cos(\frac{2j\pi}{n})} + 1) \notag\\
    c_{1,2} &= \sum^{n-1}_{j=0} (e^{2\beta_t \cos(\frac{2j\pi}{n})} - 1)
    \notag\\
    c_{1,3} &= \sum^{n-1}_{j=0} (-1)^j(e^{2\beta_t \cos(\frac{2j\pi}{n})} + 1)\notag\\
    c_{1,4} &= \sum^{n-1}_{j=0} (-1)^j(e^{2\beta_t \cos(\frac{2j\pi}{n})} - 1)
    \notag\\
    c_{2,k} & = \sum^{n-1}_{j=0} \cos(\frac{2j\pi}{n} k)e^{2\beta_t\cos(\frac{2j\pi}{n})}
\end{align}
and is inherently positive from the positivity of $T^{1}_{K}$ when $\beta_t>0$ for finite $n$. Notice that in \eq{coefficient}, the group elements with $m=1$ contribute only to the second piece in each summation of $c_r$ for $d_r = 1$ and do not contribute to $c_{2,k}$.
For the $D_4$ group considered in Sec.~\ref{sec:num}, we have
\begin{eqnarray}
\label{eq:d4eigen}
    c_{1,1} &=& 6 + e^{2\beta_t} + e^{-2\beta_t},~~c_{2,1}= e^{2\beta_t} - e^{-2\beta_t}\\\notag
    c_{1,2} &=& c_{1,3} = c_{1,4} = -2 + e^{2\beta_t} + e^{-2\beta_t}.
\end{eqnarray}
The corresponding $H_K$ after normalizing the ground state energy to zero in the character basis $|r,l\rangle$ is written as:
\begin{eqnarray}
\label{eq:D4hami}
H_K = \frac{1}{a_0} \sum_{r}\sum^{d^2_r}_{l=1}\log(\frac{c_{1,1}}{c_r}) \left|r, l\right>\left<r,l\right|
\end{eqnarray}
where $l$ represents the degree of degeneracy of the $d^2_r$ modes in the $r$ irreducible representation. The trotterization error we discussed in Sec.~\ref{sec:errorAC} would be proportional to the maximal eigenvalues of $H_K$.

Recall that $\beta_t = a/(g^2_t a_0) = 1/(g^2_H\delta_\tau)$. In the small $a_0$ region, the contribution to the $c_r$ in \eq{d4eigen} are all dominated by the $e^{2\beta_t}$ terms, which quantify the contribution from the identity group element ($j = m = 0$). Taking the limit $a_0\to 0$ for discrete groups would result in degenerate spectra:
\begin{equation}
\lim_{a_0\rightarrow 0}\frac{1}{a_0}\log\frac{c_{1,1}}{c_r} \sim\lim_{a_0\rightarrow 0}\frac{f(r)}{a_0} e^{-2\beta_t}=0
\end{equation} 
with $f(r = \{1,2\}, \{1,3\}, \{1,4\}) = 8$ and $f(r = \{2,1\}) = 6$. 
However, related to the observable $\frac{a_t m_{2\times 1}}{a_t m_{1\times 1}}$ in our study of Sec.~\ref{sec:num}, one should consider the energy ratios, say:
\begin{equation}
    \lim_{a_0\rightarrow 0}\frac{1}{a_0}\log\frac{c_{1,1}}{c_{1,2}} \bigg/\bigg(\frac{1}{a_0} \log\frac{c_{1,1}}{c_{2,1}}\bigg) = \frac{4}{3}
\end{equation}
which instead approach a finite value in the $a_0\rightarrow 0$ limit. To obtain non-degenerate spectra and have the above energy ratios fixed, one need to view $\beta_t$ and $\beta_s$ as parameters, without referring to their dependence on $a$ and $a_0$ in \eq{betas}. The continuous time limit should be taken while keeping~\cite{Harlow:2018tng}:
\begin{eqnarray}
   \lim_{a_0\rightarrow 0}\frac{1}{\beta_s}\exp(-{\beta_t}) &=& g^2_d(a) g^2_s, 
\end{eqnarray} 
where $g^2_d (a)$ is some finite constant for a given spatial lattice spacing. It then follows that:
\begin{equation}
\lim_{a_0\rightarrow 0}\frac{1}{a_0}\log\frac{c_{1,1}}{c_r} \sim\lim_{a_0\rightarrow 0}\frac{f(r)}{a_0} e^{-2\beta_t}= f(r)\frac{g^2_d(2a)}{2a }
\end{equation}
For the numerical part of our simulation in Sec.~\ref{sec:num}, we fix $\delta_t =\delta_\tau = 1$ and the corresponding dimensional spectra in the small $g^2_H$ limit is given by:
\begin{equation}
\frac{1}{c a}\log(\frac{c_{1,1}}{c_r}) \sim \frac{f(r)}{c a} \exp\bigg({-\frac{2}{g^2_H}}\bigg)
\end{equation}
which is the spectra in the above Hamiltonian limit by identifying $\exp({-\frac{2}{g^2_H(a)}})/c = g^2_d(2a)/2$. This relation also indicates that for the same spatial lattice spacing measured, the bare coupling for the the discrete group should be  exponentially suppressed than the bare coupling for a continuous group.

\section{The approximation of truncating at leading order commutators in \eq{commutator norm}}
\label{ap:perturbativity}
In this appendix, we will justify that within the region of Eq.(\ref{eq:delta_max}), the BCH operators beyond the leading order are negligible and therefore \eq{commutator norm} holds.

Define $H(k,v)$ as a commutator with $k$ powers of $\oh_K$ and $v$ powers of $\oh_V$. There are at most $4v$ links in $\oh_K$ and $2(d-1)k$ plaquettes in $\oh_V$ that can contribute to $[\oh_{K,V}, H(k,v)]$. Therefore,
\begin{align}\label{eq:H(r,s)}
    &\|[\oh_K,H(k,v)]\| \leq 8v g_H^2 \|\hat l^2 \|  \| H(k,v)\|\\
    &\|[\oh_V, H(k,v)]\|  \leq 4(d-1)k  g_H^{-2} d_U \| H(k,v)\|
\end{align}
With $r=s=1$, one has $[H_K,H_V]$ which can be bounded with $\|[A,B]\|\leq 2 \|A\| \|B\|$,
\begin{align}
 \|[\oh_K,  \oh_V]\| &\leq 2 \|\oh_K\| 2(d-1)d_Ug_H^{-2}\nonumber\\
&= \mathcal{N}_{link} 4(d-1) d_U \|\hat l^2 \|  
\end{align}
When trotterizing $U(t)$ to second order, the leading order (LO) BCH terms in $H'$ are given by $k+v=3$, with the next-to-leading (NLO) order having $k+v=5$. 
While each NLO term has a slightly different bound, they all satisfy
\begin{align}
 & \|H(k+v=5)\| \leq \notag\\
   & 4\max \left \{4(d-1) g_H^{-2} d_U,8 g_H^2 \|\hat l^2 \| \right \}^2 \|H(k+v=3)\|
   \end{align}
The coefficients $c_k$ of $H_{NLO}\equiv \delta^4\sum_k c_k\|H(k,5-k)\|$
are of $O(10^{-2})$, with the largest being $1/180$ \cite{doi:10.1063/1.3078418}. Therefore
the ratio of $H_{NLO}$ to $H_{LO}$ is
\begin{align}\label{eq:NLO ratio}
    &\frac{H_{NLO}}{H_{LO}} \lesssim 
    O(1)\frac{|\delta|^2}{(\max \delta)^2}
\end{align}
where $\max\delta=1/\max \left \{4(d-1) g_H^{-2} d_U,8 g_H^2 \|\hat l^2 \| \right \}$. When $|\delta|\leq\max\delta$ then \eq{NLO ratio} is smaller than $O(1)$. Thus, the NLO contribution is negligible and the approximation of \eq{commutator norm} is valid.
\section{The proof of the error bound for Scheme B} \label{ap:ACerror proof}
For any $\delta$ within the radius $|\delta| \leq \max \delta$, one can decompose the analytic continuation error:
\begin{align}\label{eq:lambda_f er}
    |\lambda_f(\delta, g_H^2)-\lambda(\delta, g_H^2)|\leq  |\lambda_f(\delta, g_H^2)-\lambda_f(0, g_H^2)|\notag\\ 
    + |\lambda(\delta, g_H^2)-\lambda(0, g_H^2)|+ |\lambda_f(0,g_H^2)-\lambda(0, g_H^2)|
\end{align}
The term $| \lambda(\delta, g_H^2)-\lambda(0, g_H^2)|$ is simply the trotter error, bounded by $ |\lambda(\delta, g_H^2)-\lambda(0, g_H^2)|\lesssim 
M/2$. 
We further impose a constraint $|\lambda_f(\delta, g_H^2)- \lambda_f(0, g_H^2)|\leq M/2$ on the fitting function $\lambda_f(\delta, g_H^2)$. 

 The last term in \eq{lambda_f er} is bounded by the Euclidean precision, $|\lambda_f(0,g_H^2)-\lambda(0, g_H^2)|\leq \epsilon_B \ll M$. Then there is a loose bound on $ |\lambda_f(\delta, g_H^2)-\lambda(\delta, g_H^2)|$:
\begin{align}
\label{eq:LooseBound}
    |\lambda_f(\delta, g_H^2)-\lambda(\delta, g_H^2)|\lesssim M+\epsilon_B
\end{align}
We finish the proof with the following lemma, which is a specific case of Lemma 1 in \cite{Miller:1970SIAM}.
\begin{lemma}\label{lemma:ACerror}
Let $\lambda(\delta), \lambda_f(\delta)$ be analytic in the half-disk region $\Omega=\{\delta:  \im \delta \geq 0, |\delta|\leq \max\delta\}$ with the bounds $|\lambda-\lambda_f| \leq A $ for any $\delta \in \Omega$ and $|\lambda-\lambda_f| \leq \epsilon $ on the lower boundary $\im \delta =0$. Then for any $\delta \in \Omega$, 
\begin{align}\label{eq:ACerror}
    |\lambda(\delta) -\lambda_f(\delta)| \leq \epsilon \left( \frac{A}{\epsilon}\right) ^{\re \frac{4}{\pi}\arctan \frac{\delta}{i\max\delta} }
\end{align}
\end{lemma}
\noindent\emph{Proof.}
Define
\begin{align}
    w(\delta)\equiv\frac{4}{\pi}\arctan \frac{\delta}{i\max\delta}
\end{align}
With $ \delta \in \Omega$, the range of $w$ is the infinite strip $S=\{ w: 0\leq \re w \leq 1\}$.  Define $v$ and $h$ as follows
\begin{align}
    v\equiv\ln  \frac{A}{\epsilon}
 \end{align}  
 \begin{align}\label{eq:h(delta)}
   h(\delta)\equiv\frac{e^{-vw(\delta)}}A\left[\lambda(\delta )-\lambda_f (\delta)\right]
\end{align}
The function $h(\delta)$ is analytic in the half-disk region $\Omega$. According to the maximum modulus principle, the maximum $|h(\delta)|$ can only be on the boundary of $\Omega$, which consists of two parts $\im \delta=0$ where $|\lambda_f-\lambda|\leq \epsilon$ and $|\delta|=\max \delta$ where $\re w(\delta)=1$.
\begin{align}
    |h(\delta)|_{\im \delta=0} &= \frac{1}{A}|\lambda(\delta) -\lambda_f(\delta)| _{\im \delta=0}\leq \frac{\epsilon}{A} \\
    |h(\delta)|_{| \delta|=\max\delta} &\leq |e^{-v w(\delta)}|_{| \delta|=\max\delta}\notag\\
    &=\left(\frac{\epsilon}{A}\right)^{\re w(\delta)_{| \delta|=\max\delta}}=\frac{\epsilon}{A}
\end{align}
Therefore, $|h(\delta)|\leq \frac{\epsilon}{A}$ for any $\delta\in\Omega$. Using the definition \ref{eq:h(delta)}, we obtain the following upper bound.
\begin{equation}
   |\lambda(\delta )-\lambda_f (\delta)|=A |h(\delta)|e^{v \re w(\delta)}\leq \epsilon \left (\frac{A}{\epsilon}\right)^{\re w(\delta)}
\end{equation}
For $\delta=i \delta_t$, we have $\re w(i \delta_t)= \frac{4}{\pi}\arctan \frac{\delta_t}{\max\delta}$.
In \eq{ACerror}, replace $\epsilon$  with $\epsilon_B$ and the loose upper bound $A$  with $(M+\epsilon_B)$, as \eq{LooseBound} suggests, and one gets the result of \eq{SchemeB}.

\bibliography{wise}
\end{document}